\documentclass[useAMS,usenatbib]{mn2e}
\usepackage{epsfig,amsmath,amssymb}
\usepackage{graphicx}
\usepackage{multirow}

\title[GW background from compact binary coalescences]{On the gravitational wave background from compact binary coalescences in the band of ground-based interferometers}

\author[X.-J. Zhu et al]{Xing-Jiang Zhu$^{1}$\thanks{E-mail: xingjiang.zhu@uwa.edu.au}, Eric J. Howell$^{1}$, David G. Blair$^{1}$ and Zong-Hong Zhu$^{2}$\\
$^{1}$School of Physics, University of Western
Australia, Crawley WA 6009, Australia\\
$^{2}$Department of Astronomy, Beijing Normal University, Beijing
100875, China\\}

\begin{document}

\date{}
\pagerange{\pageref{firstpage}--\pageref{lastpage}} \pubyear{2012}
\maketitle \label{firstpage}

\begin{abstract}
This paper reports a comprehensive study on the gravitational wave (GW) background from compact binary coalescences. We consider in our calculations newly available observation-based neutron star and black hole mass distributions and complete analytical waveforms that include post-Newtonian amplitude corrections. Our results show that: (i) post-Newtonian effects cause a small reduction in the GW background signal; (ii) below 100 Hz the background depends primarily on the local coalescence rate $r_0$ and the average chirp mass and is independent of the chirp mass distribution; (iii) the effects of cosmic star formation rates and delay times between the formation and merger of binaries are linear below 100 Hz and can be represented by a single parameter within a factor of $\sim 2$; (iv) a simple power law model of the energy density parameter $\Omega_{\rm{GW}}(f) \sim f ^{2/3}$ up to 50-100 Hz is sufficient to be used as a search template for ground-based interferometers. In terms of detection prospects of this background signal, we show that: (i) detection (a signal-to-noise ratio of 3) within one year of observation by the Advanced Laser Interferometer Gravitational-wave Observatory (LIGO) detectors (H1-L1) requires a coalescence rate of $r_0 = 3 \hspace{0.5mm} (0.2) \hspace{0.5mm} \rm{Mpc}^{-3} \hspace{0.5mm} \rm{Myr}^{-1}$ for binary neutron stars (binary black holes); (ii) this limit on $r_0$ could be reduced 3-fold for two co-located and co-aligned detectors, whereas the currently proposed worldwide network of advanced instruments gives only $\sim 30\%$ improvement in detectability; (iii) the improved sensitivity of the planned Einstein Telescope allows not only confident detection of the background but also the high frequency components of the spectrum to be measured, possibly enabling rate evolutionary histories and mass distributions to be probed. Finally we show that sub-threshold binary neutron star merger events produce a strong foreground, which could be an issue for future terrestrial stochastic searches of primordial GWs.
\end{abstract}


\section{Introduction}
Compact binary coalescences (CBC), of binary neutron stars (BNS), stellar mass binary black holes (BBH) and black hole-neutron stars (BH-NS), are the most promising source of gravitational waves (GWs) for ground-based interferometers such as LIGO\footnote{http://www.ligo.caltech.edu/} and Virgo\footnote{http://www.virgo.infn.it/}. Although GW detections have not been recorded so far, a few tens of detections per year should become possible when advanced detectors come online in 2015 \citep{lsc_rate10}. While individually detectable CBC events are expected within distances of hundreds of Mpc, the superposition of the gravitational radiation from these sources over cosmological volumes can form a GW background \citep[GWB;][]{Phinney01}. This signal represents another interesting target for the up-coming advanced instruments \citep[see, e.g.,][for the most recent studies]{Tania2011,BBH_Zhu11,Marassi_BBH11,Rosado11,WuCBC12,Kowalska12}.

A GWB is generally characterized by the dimensionless energy density parameter $\Omega_{\rm{GW}}(f)$, which represents the present-day fractional energy density in GWs as a function of frequency $\it{f}$. In general, assuming Newtonian energy spectra and circular binary orbits for all sources, the CBC background can be described by a power law function $\Omega_{\rm{GW}}(f) = \Omega_{\alpha} f^{\alpha}$, with $\alpha=2/3$ and an amplitude $\Omega_{\alpha}$ determined by system masses, coalescence rates and their evolution over cosmic time. Such power law models have been widely used in searches for stochastic backgrounds using LIGO/Virgo data \citep{LIGO-SGWB-limit,LSC12SGWB}, in mock data challenges \citep{Tania_ETMDC} for the third-generation detector, the planned Einstein Telescope \citep[ET;][]{et}, and in parameter estimation of a stochastic background \citep{VukPE12}.

In this paper we investigate two issues of importance for stochastic searches with ground-based interferometers. Firstly we refine the power law model for the CBC background by using complete analytical waveforms that include post-Newtonian (PN) amplitude corrections, and observation-based parameterized models of NS/BH mass distributions. The aim is to investigate what information can be extracted from a potential detection of the CBC background and to provide a ready-to-use $\Omega_{\rm{GW}}(f)$ model for CBC background searches. We secondly consider an additional motivation to study the properties of an astrophysical GWB (AGWB) -- the fact that it could act as a foreground masking the primordial GWBs from the very early Universe. As the spatial distribution of the individual sources produces time series of varying GW amplitudes, the strongest signals, which would be detected as single events, can be subtracted from the data. Therefore, as demonstrated for the BNS population using the proposed Big Bang Observer \citep{BBOsubtCutler06}, a detector with high enough sensitivity, could remove a foreground entirely by subtracting all the individually identified component signals. We show in this work that there is a significant residual foreground in the (1--500) Hz frequency range from sub-threshold BNS merger events. Such a foreground should be considered in future ground-based stochastic searches for primordial GWBs and other AGWBs.

The organization of this paper is as follows. In section \ref{AGWBgeneral} we review the theoretical framework for calculating $\Omega_{\rm{GW}}(f)$ and other quantities of an AGWB used in the literature. We also present a practical power law model of AGWBs. In section \ref{CBCsemi} we extend this model to the case of three CBC populations by considering the effects of cosmic star formation rates (CSFRs) and delay times. Then using complete waveforms we calculate semi-analytically $\Omega_{\rm{GW}}(f)$ of the CBC background. We describe in section \ref{Sim} a Monte-Carlo approach to calculate $\Omega_{\rm{GW}}(f)$ which allow NS/BH mass distributions to be included and then show how the information of mass distributions is encoded in background energy spectra. In section \ref{CalSNR} we evaluate carefully the detectability of the CBC background signal for future detectors and further investigate the construction of $\Omega_{\rm{GW}}(f)$ templates for future detectors. In section \ref{Subinds} we simulate the residual foreground noise for ET through the subtraction of the individually detectable events. In section \ref{resoveCBC} we discuss the unique time-frequency statistical properties of the CBC background and the possible implications with respect to detection. Finally we present our conclusions in section \ref{Conclusions}.
\section{Properties of an AGWB}
\label{AGWBgeneral}
In this section we summarize the broad range of formalisms used by different authors to calculate $\Omega_{\rm{GW}}(f)$ of an AGWB. We start from Phinney's practical theorem \citep{Phinney01} and compare it with various versions given in the literature. Then we derive a practical model in the general case of AGWBs. The reader who is only interested in problems related to models and the detection of the CBC background can skip this section and go straight to section \ref{CBCsemi}.

Firstly recall that $\Omega_{\rm{GW}}(f)$ is defined as the GW energy density per logarithmic frequency interval at observed frequency $\it{f}$, divided by the critical energy density required to close the Universe today $\rho_{c}=3H_{0}^{2} c^{2}/8\pi G$ with $H_{0}$ the Hubble constant. It is straightforward to compute this dimensionless function as \citep[see][for details]{Phinney01}:
\begin{equation}
\Omega_{\rm{GW}}(f)=\frac{1}{\rho_{c}}\int_{z_{\rm{min}}}^{z_{\rm{max}}} \frac{N(z)}{(1+z)} \left.\left(\frac{{\rm{d}} E_{\rm{GW}}}{{\rm{d}}\ln f_{\rm{r}}}\right)\right|_{f_{\rm{r}}=f (1+z)} {\rm{d}}z
\label{omeg},
\end{equation}
where $N(z)$ is the spatial number density of GW events at redshift z; the factor $(1+z)$ accounts for redshifting of GW energy since emission; $f_{\rm{r}}=f (1+z)$ is the GW frequency in the source frame and ${\rm{d}}E_{\rm{GW}}/{\rm{d}}\ln f_{\rm{r}}$ is the single source energy spectrum. The limits of the integral over z are given by $z_{\rm{min}}$ = max(0, $f_{\rm{r}}^{\rm{min}} / f - 1$) and $z_{\rm{max}}$ = min($z_{\ast}$, $f_{\rm{r}}^{\rm{max}} / f - 1$) with $z_{\ast}$ signifying the beginning of source formation and $f_{\rm{r}}^{\rm{min}}$ and $f_{\rm{r}}^{\rm{max}}$ for the minimum and maximal source rest-frame GW frequency respectively. Note that $f_{\rm{r}}^{\rm{min}}$, $f_{\rm{r}}^{\rm{max}}$ and ${\rm{d}}E_{\rm{GW}}/{\rm{d}}\ln f_{\rm{r}}$ depend on the source parameters (e.g., system mass) which usually follow some forms of distributions. This has been mostly neglected in previous studies and should be taken into account in order to fully characterize the background signal (it can be done through simulation as we show in section \ref{Sim}).

It is convenient to replace $N(z)$ in equation (\ref{omeg}) with the differential GW event rate ${\rm{d}}\dot{N}/{\rm{d}}z = N(z) c 4 \pi r_{z}^{2}$, where $r_{z}$ is the comoving distance related to the luminosity distance through
$d_{L}=r_{z} (1+z)$. We then obtain another version:

\begin{equation}
\Omega_{\rm{GW}}(f)=\frac{f}{\rho_{c} c}\int_{z_{\rm{min}}}^{z_{\rm{max}}} \frac{1}{4 \pi r_{z}^{2}} \left.\left(\frac{{\rm{d}}E_{\rm{GW}}}{{\rm{d}}f_{\rm{r}}}\right)\right|_{f_{\rm{r}}=f (1+z)} \frac{{\rm{d}}\dot{N}}{{\rm{d}}z} {\rm{d}}z
\label{omeg2}.
\end{equation}

The quantity given by the above integration, with units of $\rm{erg}\hspace{0.5mm}
\rm{cm}^{-2}\hspace{0.5mm} \rm{Hz}^{-1}\hspace{0.5mm} s^{-1}$, is called the \textit{spectral energy density} \citep[e.g.,][]{Ferrari99a,marassi09} or the \textit{integrated flux} \citep[e.g.,][]{Tania2011,WuCBC12}. Its dimension shows that it can be related to the specific intensity by integrating the latter over the solid angle. The first two terms inside the integral give the locally measured energy flux per unit frequency (or simply \textit{fluence}) emitted by a source at redshift z \citep{BBH98}:
\begin{equation}
\frac{{\rm{d}}E_{\rm{GW}}}{{\rm{d}}S{\rm{d}}f} = \frac{1}{4 \pi r_{z}^{2}} \left.\left(\frac{{\rm{d}}E_{\rm{GW}}}{{\rm{d}}f_{\rm{r}}}\right)\right|_{f_{\rm{r}}=f (1+z)}
\label{flux},
\end{equation}
while ${\rm{d}}\dot{N}/{\rm{d}}z$ can also be written as:
\begin{equation}
\frac{{\rm{d}}\dot{N}}{{\rm{d}}z} = \frac{R(z)}{(1+z)} \frac{{\rm{d}}V}{{\rm{d}}z} \label{dR},
\end{equation}
with the comoving volume element ${\rm{d}}V/{\rm{d}}z$ given by:
\begin{equation}
\frac{{\rm{d}}V}{{\rm{d}}z}=4\pi c\frac{r_{z}^{2}}{H(z)} \label{dVdz},
\end{equation}

\noindent where the Hubble parameter $H(z)= H_{0} \sqrt{\Omega_{\Lambda}+\Omega_{m}(1+z)^{3}}$ and $r_{z} = \int _0^z c\hspace{0.5mm} {\rm{d}}z'/H(z')$, assuming a standard $\Lambda$CDM cosmology with parameters $H_{0}=100 h\cdot \rm{km}\hspace{0.5mm} \rm{s}^{-1}\hspace{0.5mm} \rm{Mpc}^{-1}$, $h=0.7$, $\Omega_{m}=0.27$ and $\Omega_{\Lambda}=0.73$ \citep{cosmology}.

In equation (\ref{dR}) we define $R(z)=r_{0}e(z)$ \citep[see, e.g.][]{coward_GWBG_01,howell_GWBG_04}, which gives the rate density measured in cosmic time local to the event. The parameter $r_{0}$ is the local rate density, usually used to estimate detection rates for different detectors, and $e(z)$ is a dimensionless factor which models the source rate evolution over cosmic time. The later is usually associated with the CSFR for stellar catastrophic GW events.

The factor $(1+z)$ in equation (\ref{dR}) converts $R(z)$ to an earth time based quantity. The statement that such a factor does not exist given in \citet{araujo05} does not change the calculation of $\Omega_{\rm{GW}}(f)$ as the factor appears additionally in their equation for ${\rm{d}}E/{\rm{d}}S{\rm{d}}f$. This caveat has also appeared in other publications, e.g., \citet{regimbau08,CCSN-limit,r-mode2011,BBH_Zhu11,eric2011}. We correct it with equations (\ref{flux}) and (\ref{dR}) since they provide physically correct estimates of the corresponding quantities.

Combining equations (\ref{omeg2})-(\ref{dVdz}) yields the compact form:
\begin{equation}
\Omega_{\rm{GW}}(f)=\frac{f}{\rho_{c}} \frac{r_{0}}{H_{0}} \int_{z_{\rm{min}}}^{z_{\rm{max}}} \hspace{-2mm} \frac{e(z)}{(1+z)\sqrt{\Omega_{\Lambda}+\Omega_{m}(1+z)^{3}}} \frac{{\rm{d}}E_{\rm{GW}}}{{\rm{d}}f} {\rm{d}}z
\label{omeg3}.
\end{equation}

Alternatively $\Omega_{\rm{GW}}(f)$ can be calculated through the single-source characteristic amplitude $h_{c}(f) = f\langle|\tilde{h}(f)|\rangle$ with $\langle|\tilde{h}(f)|\rangle$ denoting the frequency-domain GW amplitude (in Hz$^{-1}$) averaged over source orientations. In this case one can use the following relation to replace equation (\ref{flux}):
\begin{equation}
\frac{{\rm{d}}E_{\rm{GW}}}{{\rm{d}}S{\rm{d}}f} = \frac{\pi c^{3}}{2 G} h_{c}^{2}(f)
\label{flux2}.
\end{equation}

The average over all source orientations for an inspiraling binary, a rotating NS or a ringing BH is given by \citep{lrr-2009-2}:
\begin{equation}
\int_{-1}^{1} {\rm{d}}(\cos \iota) \left[\left(\frac{1+\cos ^{2}\iota}{2}\right)^{2} + \cos ^{2}\iota \right] = \frac{4}{5}
\label{average1},
\end{equation}
where $\iota$ is the inclination angle of the characteristic direction of the source, determined by the orbital or spin angular momentum, with respect to the line of sight. This shows that $\langle|\tilde{h}(f)|\rangle$ is smaller than that of an optimally oriented source (i.e., $\iota=0$) by a factor of $4/5$.

The one-sided spectral density of a GWB, $S_{h}(f)$, can be conveniently compared with detector sensitivities and is related to $\Omega_{\rm{GW}}(f)$ through \citep{magg00}:
\begin{equation}
S_{h}(f) = \frac{3 H_{0}^{2}}{2 \pi^{2}} f^{-3} \Omega_{\rm{GW}}(f)
\label{Sh}.
\end{equation}
Note that assuming an isotropic GWB, a factor of $1/5$ should be included to account for the average detector response over all source locations in the sky, when the above equation is used directly to compare $S_{h}(f)$ with noise power spectral densities of L-shaped interferometers. For instruments with non-perpendicular arms, it becomes $\sin^{2} \zeta/5$ with $\zeta$ being the opening angle between the two arms.

Another important quantity of an AGWB is the (dimensionless) duty cycle, $\xi$, which describes the degree of overlap of individual signals in time domain. It can be computed as \citep[see, e.g.,][]{coward_regimbau_06}:
\begin{equation}
\xi =\int_{0}^{z_{\ast}} \Delta \tau \frac{{\rm{d}}\dot{N}}{{\rm{d}}z} {\rm{d}}z \label{DC},
\end{equation}
where $\Delta \tau$ is the average observed signal duration. A value of $\xi \geqslant 1$ generally implies a continuous background. Here $\Delta \tau$ is assumed to be frequency independent; If a dependence exists, the upper limit of the integration should be changed to $z_{\rm{max}}$. The above defined duty cycle may not be useful if there is significant frequency evolution of $\Delta \tau$, e.g., for CBC sources -- this will be further discussed in section \ref{resoveCBC}.

\subsection{A practical model}
\label{pmodel}
We now derive a practical model for AGWBs formed by sources for which the gravitational energy spectrum can be approximated by a power law function of frequency. Such a case is of particular interest because $\Omega_{\rm{GW}}(f) \sim f^{\alpha}$ is naturally obtained when ${\rm{d}}E_{\rm{GW}}/{\rm{d}}f_{\rm{r}} = A f_{\rm{r}}^{\alpha-1}$ with $A$ the overall amplitude. Then equation (\ref{omeg3}) has a simple form within the frequency range $f_{\rm{r}}^{\rm{min}} \leqslant f \leqslant f_{\rm{r}}^{\rm{max}}/(1+z_{\ast})$:
\begin{equation}
\Omega_{\rm{GW}}(f, \alpha) = \frac{A}{\rho_{c}}\, \frac{r_{0}}{H_{0}}\, f^{\alpha}\, J(\alpha)
\label{Omgw1},
\end{equation}
where we have defined a dimensionless function:
\begin{equation}
J(\alpha) = \int_{0}^{z_{\ast}} \frac{e(z)(1+z)^{\alpha-2}}{\sqrt{\Omega_{\Lambda}+\Omega_{m}(1+z)^{3}}}\, {\rm{d}}z
\label{Intz}.
\end{equation}
Note that we have changed the lower and upper limit of the integral from $z_{\rm{min}}$ and $z_{\rm{max}}$ to 0 and $z_{\ast}$ respectively, as we focus on the particular frequency range $f_{\rm{r}}^{\rm{min}} \leqslant f \leqslant f_{\rm{r}}^{\rm{max}}/(1+z_{\ast})$ where the power law relation applies.

We define $e(z)= \dot{\rho}_{\ast}(z) / \dot{\rho}_{\ast}(0)$, where $\dot{\rho}_{\ast}(z)$ is the CSFR density (in $M_{\odot} \, \rm{yr}^{-1} \, \rm{Mpc}^{-3}$). The assumption made here is that the GW event rate closely tracks the CSFR, e.g., in core collapse supernovae related mechanisms; Otherwise effects of delay times should be included as we show in section \ref{smodelCBC} for CBC events. Note that $r_{0}$ is equivalent to the parameter $\lambda$ used in some studies to represent the fraction of stellar mass converted to GW source progenitors \citep[see, e.g., ][]{regimbau08,WuCBC12}. As estimates of $r_{0}$ do not normally rely on measurements of $\dot{\rho}_{\ast}(0)$; rather they can be based on independent observations or theoretical calculations, we choose to treat $r_{0}$ as a free parameter, independent on the CSFR models throughout the paper.

\begin{figure}
\includegraphics[width=0.48\textwidth]{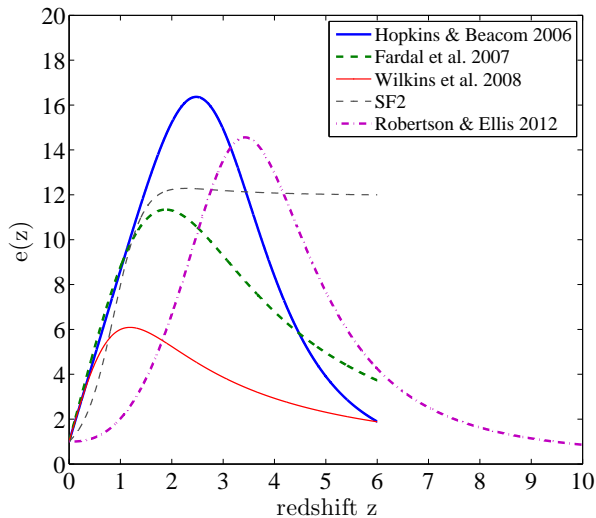}
\caption{The dimensionless rate evolution factor $e(z)$ based on different parameterized models of CSFR. SF2 is taken from \citet{Porciani_3SFR01}, and we use the ``low rate'' gamma ray bursts derived model of \citet{SFR_GRB11}.}
\label{ezSFR}
\end{figure}

In this work we consider five parameterized forms of $\dot{\rho}_{\ast}(z)$ derived from various observations \citep[see][for details]{Porciani_3SFR01,HB06,SFR_Fardal07,SFR_Wilkins08,SFR_GRB11}. The corresponding models of $e(z)$ are shown in Figure \ref{ezSFR}. We therefore set $z_{\ast}$ as the maximal redshift for which the CSFR model is applicable: $z_{\ast}=15$ for the recent study of \citet{SFR_GRB11} which is derived from gamma ray burst observations and $z_{\ast}=6$ for the other four models.

\begin{figure}
\includegraphics[width=0.46\textwidth]{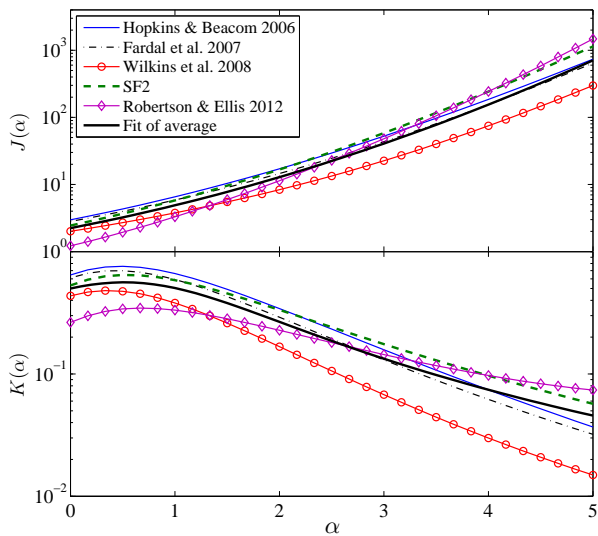}
\caption{The dimensionless function $J(\alpha)$ (upper panel) given by equation (\ref{Intz}) and $K(\alpha)$ (lower panel) defined by equation (\ref{Kalpha}) calculated for the five models of $e(z)$ shown in Figure \ref{ezSFR}. The bold line is a fit of average over the five models.}
\label{ConstAlpha}
\end{figure}

Figure \ref{ConstAlpha} (upper panel) shows $J(\alpha)$ calculated for $\alpha = [0, \hspace{0.5mm}5]$ using the five models of $e(z)$. Since all the current predictions of AGWBs in the frequency band of terrestrial detectors indicate that $\Omega_{\rm{GW}}(f)$ increases from about 10 Hz to several hundreds Hz \citep[see, e.g., Figure 6 of][]{Tania2011}, the chosen range of $\alpha$ is adequate for most of possible scenarios, e.g., $\alpha=2/3$ for inspiraling compact binaries as mentioned earlier, $\alpha=2$ for NS r-mode instabilities \citep{r-mode98,r-mode99,r-mode2011}, and $\alpha=4$ for magnetars \citep{Regimbau06,magnetar11}. The five curves of $J(\alpha)$ are within a factor of 2 around the average, for which a least-square fit is $\log[J(\alpha)] = 0.04 \alpha^{2} + 0.3\alpha +0.35$.

For a power law energy spectrum, the total GW energy emitted in the frequency range $(f_{\rm{r}}^{\rm{min}}, \hspace{0.5mm} f_{\rm{r}}^{\rm{max}})$ is $\Delta E_{\rm{GW}} = A \int f^{\alpha-1} {\rm{d}}f$. We further define a dimensionless function:
\begin{equation}
K(\alpha) = A\, ({100\hspace{0.5mm} \rm{Hz}})^{\alpha} \frac{J(\alpha)}{\Delta E_{\rm{GW}}}
\label{Kalpha},
\end{equation}
to obtain a practical form:
\begin{eqnarray}
\Omega_{\rm{GW}}(f, \alpha)  &=& 10^{-9} \, \left(\frac{r_0}{1\, \rm{Mpc}^{-3}\rm{Myr}^{-1}}\right)\, \left(\frac{\Delta E_{\rm{GW}}}{0.01\, M_{\odot} c^2}\right)  \nonumber\\&& \left(\frac{f}{100 \rm{Hz}}\right)^{\alpha} K(\alpha)\;.
\label{Omgw0}
\end{eqnarray}

The function $K(\alpha)$ obtained while arbitrarily setting $f_{\rm{r}}^{\rm{min}}=$10 Hz and $f_{\rm{r}}^{\rm{max}}=$1000 Hz is shown in the lower panel of Figure \ref{ConstAlpha}. The least-square fit of the average over the five models of $e(z)$ is given by $K(\alpha) = (1.2-0.04\alpha)/(\alpha^2 -1.1\alpha+2.4)$. Note that the chosen values of $(f_{\rm{r}}^{\rm{min}}, \hspace{0.5mm} f_{\rm{r}}^{\rm{max}})$ correspond to a frequency band where ground-based detectors have significant sensitivities; One can calculate $K(\alpha)$ for a specific type of source using equation (\ref{Kalpha}). Figure \ref{ConstAlpha} implies that: a) as $\alpha$ increases, the high redshift ($z \gtrsim 4$) sources contribute more to the background; b) effects of the CSFR introduce uncertainties in the overall amplitudes of $\Omega_{\rm{GW}}(f)$ within a factor of about 2 for $\alpha \lesssim 3$ and up to 5 for larger $\alpha$.

Combining equations (\ref{Omgw0}) and (\ref{Sh}) gives $S_{h}(f)$ in a convenient form:
\begin{eqnarray}
S_{h}^{1/2}(f, \alpha)\hspace{-2mm} &=&\hspace{-2mm} 1.3 \times10^{-26}\, \mbox{Hz}^{-{\frac{1}{2}}} \left(\frac{f}{100\hspace{0.5mm} \rm{Hz}}\right)^{\frac{(\alpha-3)}{2}}  [K(\alpha)]^{\frac{1}{2}} \nonumber\\&& \hspace{-2mm} \left(\frac{r_0}{1\hspace{0.5mm} \rm{Mpc}^{-3}\rm{Myr}^{-1}}\right)^{\frac{1}{2}} \left(\frac{\Delta E_{\rm{GW}}}{0.01\hspace{0.5mm} M_{\odot} c^2}\right)^{\frac{1}{2}}
\label{Sh0}.
\end{eqnarray}

Similarly, a convenient relation between the duty cycle $\xi$ of an AGWB and $\Delta \tau$ and $r_0$ can be obtained by combining equations (\ref{dR}), (\ref{dVdz}) and (\ref{DC}) and averaging over the five models of $e(z)$ shown in Figure \ref{ezSFR}:
\begin{equation}
\xi = 0.2\, \left(\frac{\Delta \tau}{ 1\, \rm{sec}}\right)\, \left(\frac{r_0}{1\hspace{0.5mm} \rm{Mpc}^{-3}\rm{Myr}^{-1}}\right)
\label{DC0}.
\end{equation}
Such a relation shows if there is a continuous GWB formed by one particular type of sources.

Equations (\ref{Omgw1})-(\ref{Sh0}) represent our practical power law model for AGWBs. The power law relation holds for the frequency range $[f_{\rm{r}}^{\rm{min}}, \hspace{0.5mm} f_{\rm{r}}^{\rm{max}}/(1+z_{\ast})]$, where effects due to rate evolutionary histories are linear. The model allows quick evaluation of the background signal strength and its uncertainty using estimates of $r_0$ and $\Delta E_{\rm{GW}}$ (which are also essential for back-of-the-envelope predictions of single-source detection prospects). As our knowledge improves the model can be easily modified to provide templates for future stochastic background searches. In the following sections we will develop a ready-to-use model for the CBC background by considering additional issues that have not been considered here.

\section{The CBC background: analytical approaches}
\label{CBCsemi}
In this section we extend the derivation in subsection \ref{pmodel}, to obtain models for CBC events analytically.

\subsection{A simple power law model}
\label{smodelCBC}
Previous calculations of $\Omega_{\rm{GW}}(f)$ for the CBC background have employed the Newtonian inspiral energy spectrum, with the exception of BBH \citep{BBH_Zhu11,Marassi_BBH11,WuCBC12}. Following the previous derivation, we present here a simple power law model generalized for three CBC populations.

In the Newtonian limit, the GW energy spectrum for an inspiralling circular binary of component masses $m_1$ and $m_2$ is given by \citep[see, e.g.,][]{thorne87}:
\begin{equation}
\frac{{\rm{d}}E_{\rm{GW}}}{{\rm{d}}f_{\rm{r}}} = \frac{(\pi G)^{2/3} M_c^{5/3}}{3} f_{\rm{r}}^{-1/3} ,
\label{dEdf}
\end{equation}
where $M_c$ is the chirp mass defined as $M_c = M \eta ^{5/3}$, with $M= m_1+m_2$ the total mass and $\eta = m_1 m_2 /M^{2}$ the symmetric mass ratio. Inserting this into equation (\ref{omeg3}) and combining the expression of $\rho_{c}$ gives ($f_{\rm{r}}^{\rm{min}} \leqslant f \leqslant f_{\rm{r}}^{\rm{max}}/(1+z_{\ast})$):
\begin{equation}
\Omega_{\rm{GW}}(f)=\frac{8}{9} \frac{1}{c^2 H_{0}^{2}} \frac{r_{0}}{H_{0}} (\pi G M_c)^{5/3} f^{2/3} J_{2/3}
\label{omegCBC},
\end{equation}
where we have defined a dimensionless quantity:
\begin{equation}
J_{2/3} = \int_{0}^{z_{\ast}} \frac{e(z)(1+z)^{-4/3}}{\sqrt{\Omega_{\Lambda}+\Omega_{m}(1+z)^{3}}} {\rm{d}}z
\label{J23CBC}.
\end{equation}
To determine the applicable frequency range of the above power law relation, one has $f_{\rm{r}}^{\rm{min}}$ well below 1 Hz and $f_{\rm{r}}^{\rm{max}}$ given by the frequency at the last stable orbit (LSO) during inspiral $f_{\rm{LSO}} \simeq 4400\hspace{0.5mm} {\rm{Hz}}/M$ with $M$ in units of $M_{\odot}$.

The newly defined quantity $J_{2/3}$ differs from $J(2/3)$ as given in equation (\ref{Intz}) in the definition of $e(z)$: for CBC events, effects due to the delay time $t_{d}$ between the formation and the final merger of binaries should be taken into account. By assuming compact binary formation closely tracks the cosmic star formation, we define $e(z)= \dot{\rho}_{\ast,c}(z)/\dot{\rho}_{\ast,c}(0)$ by introducing a $\dot{\rho}_{\ast}$-related quantity:
\begin{equation}
\dot{\rho}_{\ast,c}(z)= \int_{t_{\rm{min}}}^{t_{\ast}} \dot{\rho}_{\ast}(z_{f}) \frac{{\rm{d}}t_{f}}{{\rm{d}}t_{z}} P(t_{d}) {\rm{d}}t_{d},
\label{rhos}
\end{equation}
\noindent where $P(t_{d})$ and $t_{\rm{min}}$ denote the probability distribution for and minimum value of $t_{d}$ respectively. The upper limit of the integral $t_{\ast}$ corresponds to $z_{\ast}$. For CBC events, $P(t_{d})$ follows a $1/t_{d}$ form\footnote{We note that $P(t_{d})$ of type Ia supernovae was observationally found to be consistent with the $1/t_{d}$ predictions for progenitors of the GW induced binary white dwarf mergers \citep{SNtdObs11,Maoz12SNIaDTD}.} as suggested by latest population-synthesis studies on compact binary evolution \citep{Dominik12StarTrack}. The parameters $z$ and $z_f$ are the redshifts when a GW event occurred and the system was initially formed respectively, with corresponding time coordinates $t_{z}$ and $t_{f}$. In our fiducial cosmology, $t_{d}$ is given by the lookback time between $z$ and $z_f$, integrating ${\rm{d}}z'/[(1+z')H(z')]$ from $z$ to $z_f$. The term ${\rm{d}}t_{f}/{\rm{d}}t_{z}=(1+z)/(1+z_{f})$ is included to convert a rate at $t_{f}$ ($\dot{\rho}_{\ast}(z_{f})$) to one local to $t_{z}$ ($\dot{\rho}_{\ast,s}(z)$).

It should be mentioned that our equation (\ref{rhos}) is equivalent to equation (2) of \citet{Regimbau_Hughes09} by noting that the $(1+z)$ factor cancels with the one in equation (\ref{dR}). The additional $(1+z)$ term in equation (9) of \citet{BBH_Zhu11} was an error, and lead to a factor of 2 underestimate of the BBH background signal.

For the five considered CSFR models and for a minimum delay time $t_{\rm{min}}$ in the range of $10-100$ Myr, $J_{2/3}$ is well constrained within $(1.3 - 2.6)$. It is roughly a factor of 2 smaller than $J(2/3)$ as given by equation (\ref{Intz}) and shown in Figure \ref{ConstAlpha} where no time delay is assumed. Our selected range of $t_{\rm{min}}$ is largely consistent with results presented in \citet{Dominik12StarTrack}; see their Figures 14-17 for details. We note, however, that in some extreme cases $t_{\rm{min}}$ for BBHs could be much higher, e.g., 500 Myr. This does not change our results significantly as we will show below. For the commonly used CSFR of \citet{HB06}, $J_{2/3}$ as a function of $t_{\rm{min}}$ (in Myr) can be expressed as:
\begin{equation}
J_{2/3}(t_{\rm{min}}) = 3.67-0.85\hspace{0.5mm} (t_{\rm{min}})^{0.165}
\label{J23HB06},
\end{equation}
for $10 {\rm{Myr}} \leqslant t_{\rm{min}} \leqslant 500 {\rm{Myr}}$; increasing $t_{\rm{min}}$ from 100 Myr to 500 Myr reduces $J_{2/3}$ from 1.85 to 1.3.

Replacing the constants with their numerical values, equation (\ref{omegCBC}) becomes:
\begin{eqnarray}
\Omega_{\rm{GW}}(f) \hspace{-1mm}  &=& \hspace{-1mm}  9.1 \times 10^{-10} \left(\frac{r_0}{1\hspace{0.5mm}\rm{Mpc}^{-3}\rm{Myr}^{-1}}\right) \left(\frac{\langle M_c^{5/3} \rangle}{1\hspace{0.5mm} M_{\odot}^{5/3}}\right) \nonumber\\&& \hspace{-2mm} \times \frac{J_{2/3}}{2} \left(f \over {100\hspace{0.5mm} \rm{Hz}}\right) ^{2/3}
\label{omegCBC0}.
\end{eqnarray}
Here we have replaced $M_c^{5/3}$ in equation (\ref{omegCBC}) with $\langle M_c^{5/3} \rangle$ to account for a distribution of system masses -- the consideration of $\langle M_c^{5/3} \rangle$ other than $\langle M_c \rangle ^{5/3}$ is based on the fact that $\Omega_{\rm{GW}}(f)$ is an average over individual energy spectra characterized by $M_c^{5/3}$. As the differences between the two quantities are very small (as we will show in Table \ref{tb2}), we do not attempt to distinguish between them and will use the term \textit{average chirp mass}. Note that the CBC background signal contains information about the physical chirp mass, while single event detections normally measure the redshifted chirp mass $M_c (1+z)$ \citep{CutlerFlanagan94}.

We have reviewed calculations of $\Omega_{\rm{GW}}(f)$ for the CBC background through a simple power law model. The model extends that of \citet{Phinney01} by considering different rate evolutionary histories and combining uncertainties associated with CSFRs and delay times into a single parameter $J_{2/3}$. In the next subsection we will introduce some additional inputs to produce more accurate estimates.
\subsection{Beyond a simple power law}
We consider for the first time new information of two aspects to refine previous estimates:\newline
1) Observation-based parameterized models of NS/BH mass distribution -- through a Monte-Carlo simulation (section \ref{Sim}), we will in subsection \ref{NumResults} investigate how the spectral shape of the background depends on the mass distributions;\newline
2) Up-to-date complete waveforms for populations of BNS, BBH and BH-NS systems -- these will show how well a CBC background can be approximated by a simple power law model in the ground-based detector band.\newline
The main parameters are summarized in Tables \ref{tb1} and \ref{tb2}; for the interested readers we provide an overview below.  Unless we otherwise specify, we will use the information contained in Table \ref{tb1} and \ref{tb2}, and the CSFR of \citet{HB06} in the following sections.

\subsubsection{Observational inputs}
We consider the parameterized models of NS/BH mass distribution recently derived from observational mass measurements. For NSs that are observed in double NS systems (with one or two pulsars), high-precision mass measurements are available \citep[see Table 1 in][and references therein]{NSmass1201}, indicating a very narrow distribution. Using the observational data for the 6 double NS systems, \citet{NSmass1201} found that the NS mass distribution can be well described by a Gaussian with a mean $\mu=1.33 M_{\odot}$ and a standard deviation $\sigma=0.06 M_{\odot}$.

In contrast with the consensus on the narrowness of the NS mass distribution, the BH mass measurements are subject to much larger uncertainties, leading to a greater range in inferred distribution. Utilizing the maximal amount of observational information available for 16 BHs in transient low-mass X-ray binaries, \citet{BHmass10} concluded that the underlying mass distribution can be best described by a Gaussian with $\mu=7.8 M_{\odot}$ and $\sigma=1.2 M_{\odot}$. More recently, \citet{FarrBHmass11} considered a broad range of parameterized models, and using a Bayesian model selection analysis, they found a Gaussian and a power law distribution are preferred for low-mass X-ray binaries, whereas an exponential distribution and a two-Gaussian model are favored if 5 high-mass, wind-fed X-ray binary systems were included \citep[see][for details]{FarrBHmass11}.

Unless stated explicitly in Table \ref{tb1} our considered mass (in $M_{\odot}$) interval for NS (BH) is $[1,\hspace{0.5mm} 2] \hspace{1mm} ([4,\hspace{0.5mm} 40])$. Given the adopted models of distributions, it is highly unlikely to obtain masses outside these intervals. We note that the existence of a ``gap'' between the maximum NS mass and the lower bound of observationally inferred BH masses has been suggested in \citet{BHmass10} and \citet{FarrBHmass11}. Such a ``gap'' can not be attributed to observational selection effects as concluded in the former paper. In this regard, terrestrial advanced GW detectors will be able to resolve this problem through precise measurements of NS/BH masses from tens up to hundreds of detections of CBC events.

The BH spin distribution is highly uncertain - currently there have been only about 10 stellar mass BHs with (model dependent) spin estimates available \citep{BHspin09,BHspin11}. Considering recent results on the determination of the extreme spin of the BH in Cygnus X--1 \citep{GouBHspin11,FabianBHspin12}, we assume a uniform distribution with spin parameter $\chi=S_{a}/m^2$ between -0.95 and 0.95, where $S_{a}$ is the spin angular momentum and $m$ is the BH mass and positive or negative value of $\chi$ implies alignment or anti-alignment between component spin and orbital angular momentum. As most NSs are observed to be weakly spinning \citep{ATNF05Pulsar}, and the fastest spinning NS in double pulsar systems, PSR J0737–-3039A, has a spin period of 22.70 ms \citep{Nature03NS} and equivalently $\chi \sim 0.05$ \citep{Brown12NSaLIGO}, we neglect the spin of NSs in our analysis.

Observational NS/BH mass measurements were also used as inputs or calibrations in the population-synthesis simulations adopted by \citet{Marassi_BBH11} and \citet{Kowalska12}. Results of these studies are based on chirp mass distributions of some simulated populations of CBC sources. We assume in this paper that components of coalescing compact binaries follow the observational mass/spin distributions. Note that: a) for BNS, simulated chirp mass distribution presented in \citet{Dominik12StarTrack} is also very narrow and should give similar results to what we will obtain in the following sections; b) Our adopted BH mass/spin distributions only apply to BHs in X-ray binaries and may not be representative for BBH and BH-NS systems.

\begin{table}
\begin{center}
\caption{\label{tb1} NS/BH mass ($m$) and spin ($\chi$) distribution.}
\vspace{-2mm}
\begin{tabular}{lcccc}
  \hline
  \hline
         & $\chi$             &  $m$     & $\langle m \rangle$  & Ref   \\
  \hline
   NS    & ...          & $N(1.33,0.06)$ & 1.33           &    (1)    \\
  \hline
   \multirow{4}{*}{BH}    & \multirow{4}{*}{$U(-0.95,0.95)$}  & $N(7.8,1.2)$   &  7.8   &   (2)     \\
       &                    &  Power law    & 7.35           &     \multirow{3}{*}{(3)}   \\
       &                    &  Exponential   & 10             &        \\
       &                    &  Two-Gaussian  &  10            &        \\
  \hline
  \hline
\end{tabular}
\end{center}
\vspace{-3mm}
\medskip{Notes: All values of mass are in $M_{\odot}$. $N(\mu,\sigma)$ implies a Gaussian distribution with a mean $\mu$ and a standard deviation $\sigma$; $U(a,b)$ is a uniform distribution between $a$ and $b$. The upper/lower bound of BH spin corresponds to the recently determined extreme spin of the BH in Cygnus X--1 \citep{GouBHspin11,FabianBHspin12}; positive or negative $\chi$ implies alignment or anti-alignment between component spin and orbital angular momentum. References for mass distributions: (1) \citet{NSmass1201}; (2) \citet{BHmass10} -- Gaussian BH mass, which is used as our fiducial model; (3) \citet{FarrBHmass11} -- for the other three models of BH mass distribution, and we use the median values of parameters given in the paper: Power law -- $P(m) \sim m^{-6.4}$ for $6 \leqslant m \leqslant 23$; Exponential -- $P(m) \sim e^{m/m_0}$, with $m_0=4.7$ for $m \geqslant 5.33$; Two-Gaussian -- $N(7.5,1.3)$ and $N(20.4,4.4)$ with weights 0.8 and 0.2 respectively. The power law model in Ref (3) has a slightly lower mean $\mu=7.35 M_{\odot}$ due to the exclusion of one low-mass X-ray binary system in their analysis as compared to Ref (2).}
\end{table}

\begin{table}
\begin{center}
\caption{\label{tb2} Information about CBC populations used in this work.}
\vspace{-1.5mm}
\begin{tabular}{lcccc}
  \hline
  \hline
       & \multirow{2}{*}{waveform} & $r_0$  & $t_{\rm{min}}$ &  $\langle M_c^{5/3} \rangle$ \\
  \vspace{-1mm}
       & & ($\rm{Mpc}^{-3} \hspace{0.5mm} \rm{Myr}^{-1}$) & (Myr) & ($M_{\odot}^{5/3}$) \\
  \hline
   BNS  & TaylorT4  & 1     & 20 &  1.276   \\
  \hline
\multirow{4}{*}{BH-NS} & \multirow{4}{*}{IMR} & \multirow{4}{*}{0.03}  & \multirow{4}{*}{30} &  4.948  \\
    &       &    &    &   4.734   \\
         &       &    &    &   5.795   \\
         &       &    &    &  5.779    \\
  \hline
  \multirow{4}{*}{BBH}      & \multirow{4}{*}{IMR} & \multirow{4}{*}{0.005} & \multirow{4}{*}{50} &  24.22  \\
      &       &    &    &  21.86 \\
         &       &    &    &  35.40 \\
         &       &    &   &  35.29  \\
  \hline
  \hline
\end{tabular}
\end{center}
\vspace{-3mm}
\medskip{Notes: IMR -- the phenomenological inspiral-merger-ringdown waveform for non-precessing spinning BBHs presented in \citet{IMR_Ajith11}; we also use this model for BH-NS as an approximation to the type-II spectrum found in numerical simulations \citep{BH-NS_review11}. For BNS waveform we adopt the TaylorT4 formula with 3.0 PN amplitude accuracy given in \citet{Blanchet3PN08}. Values of $r_0$ correspond to the realistic estimates in \citet{lsc_rate10}. $t_{\rm{min}}$ given here is used as the fiducial value, based on the standard Submodel A for solar metallicity $Z_{\odot}$ in \citet{Dominik12StarTrack} -- see Figure 8 therein; we also consider a range of 10-100 Myr to account for uncertainties. The quantities $\langle M_c^{5/3} \rangle$ are calculated using mass distributions presented in Table \ref{tb1} and assuming component masses are uncorrelated and follow the same distribution for BNS and BBH; four values for BH-NS and BBH are given in order from top to bottom as for a Gaussian, Power law, Exponential and Two-Gaussian BH mass distribution. We note that the quantity $\langle M_c \rangle ^{5/3}$ is smaller than $\langle M_c^{5/3} \rangle$ by $< 1\%$ for BNS and the first two entries of BH-NS and BBH, and about $2\%$ ($4\%$) for the other BH-NS (BBH) values -- we go with the latter quantity throughout the paper, but also use the former when comparing with other studies (in which case we neglect their differences).}
\end{table}

\subsubsection{Up-to-date analytical complete waveforms}
\label{Newwaveforms}
Following \citet{BBH_Zhu11}, we use the phenomenological inspiral-merger-ringdown waveforms for non-precessing spinning BBHs presented in \citet{IMR_Ajith11}. In this model the TaylorT1 waveform is adopted for the inspiral phase, with 1.5 PN order amplitude corrections to the Newtonian waveform \citep{Arunsp15PN}. We note that the waveform model is calibrated against numerical relativity simulations in the parameter range of mass ratios between 1 and 4 and $\chi$ between -0.85 and 0.85, but we employ it for slightly broader parameter space. Our calculations can be improved once more accurate and general models become available.

As no phenomenological complete waveforms are currently available for BNS\footnote{Work is in preparation by the numerical relativity group at AEI and collaborators (L. Rezzolla 2012, private communication).} and BH-NS systems, we consider analytical models that approximate the waveforms given by numerical relativity simulations. For BH-NS, we use the same model as that for BBH. The justification for our choice is two-fold. Firstly, the type-II spectrum found in numerical simulations is similar to that of a BBH with the same mass ratio \citep{BHNS09,BH-NS_review11}, showing a clear signature of inspiral, merger and ringdown. This happens primarily for larger mass ratios ($\gtrsim 3-5$) when the smaller NS is simply swallowed by the BH. For the NS/BH mass distribution used in this work this condition is largely fulfilled. Secondly, PN amplitude corrections and effects of BH spins can be included by using the adopted BBH model.

For BNS, we use the TaylorT4 point-particle waveform with 3.0 PN order amplitude accuracy \citep{Blanchet3PN08}. We apply the waveform up to 5000 Hz to account for a realistic cutoff of the complete spectrum. Comparisons between the TaylorT4 waveform and numerical relativity results generally indicate that the former underestimates the post-merger emission \citep{BNS09044551,BNSreview12}. Recently \citet{BNSsim12} found that the generic outcome of two $1.35 M_{\odot}$ NS mergers is the formation of a deformed differentially rotating massive NS, and that violent oscillations of the merger remnants lead to a pronounced peak in the GW spectra. We note that the peaks shown in this work are sharper than results obtained in full general relativistic simulations \citep[see, e.g.,][]{BNS09044551,Luciano10BNS}, largely due to a different numerical treatment.

\begin{figure}
\includegraphics[width=0.48\textwidth]{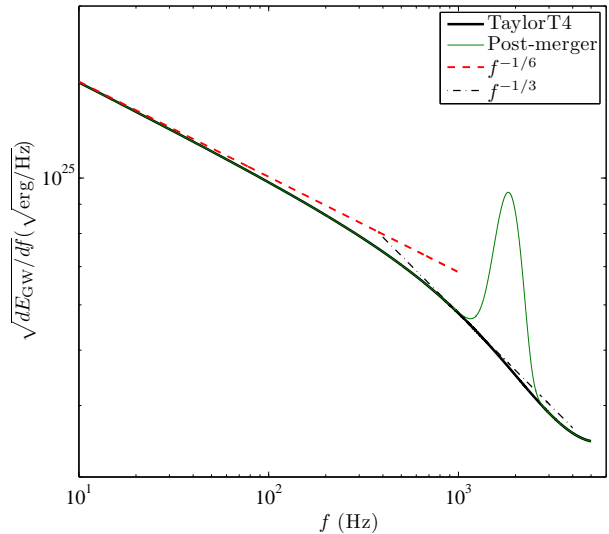}
\caption{The GW energy spectrum for a BNS of equal mass $1.33 M_{\odot}$ calculated using the TaylorT4 formula. The post-merger signal is represented (optimistically) by a Gaussian spectrum centered at around 2 kHz. Also shown are curves of $f^{-1/6}$ and $f^{-1/3}$ -- $f^{-1/6}$ corresponds to the Newtonian inspiral spectrum given by equation (\ref{dEdf}) and $f^{-1/3}$ shows the gradient at around 1 kHz. The curves are displayed as the square root of ${\rm{d}}E_{\rm{GW}}/{\rm{d}}f$ in order to be directly comparable to the quantity $h_{\rm{eff}}=f|\tilde{h}(f)|$ commonly used in the numerical relativity community.}
\label{FigdedfBNS}
\end{figure}

Figure \ref{FigdedfBNS} shows the energy spectra for a BNS of equal mass $1.33 M_{\odot}$. We consider a simple Gaussian spectrum to investigate the possible contribution from the post-merger emission to the GWB. Following \citet{CCSN-limit}, we take the form of ${\rm{d}}E_{\rm{GW}}/{\rm{d}}f=A \exp [-(f-f_{\rm{peak}})^2 / 2 \Delta^2]$ where $A$ arbitrarily set to be twice that of TaylorT4 waveform at 1000 Hz, $f_{\rm{peak}} = 1840$ Hz and $\Delta =250$ Hz -- $f_{\rm{peak}}$ corresponds to the lowest value given in Table 2 of \citet{BNSsim12} and we use a much higher width $\Delta$. The chosen parameters give a optimistic representation of post-merger emission because: a) depending on NS equation of state, the peak frequency can be higher (up to about 4 kHz), together with narrower peaks, making it harder to detect (in terms of both single events and the contribution to a GWB); b) for the case of prompt BH formation (mainly for larger binary masses) the peaks are much smaller. Due to these uncertainties, the above mentioned Gaussian spectrum will be used only for semi-analytical calculations presented in the next subsection.

\subsection{Semi-analytical results}
\label{SemiModels}
\begin{figure}
\includegraphics[width=0.48\textwidth]{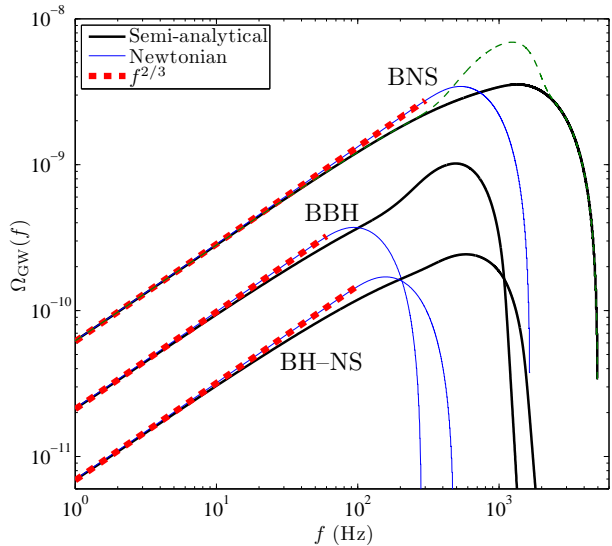}
\caption{The energy density parameter $\Omega_{\rm{GW}}(f)$ of the GWBs formed by three CBC populations (BNS, BH-NS, and BBH) calculated using complete waveforms (semi-analytical), compared with Newtonian models and simple power law models $f^{2/3}$ (see text). For BNS, the bump at around 1 kHz corresponds to the (optimistic) contribution from the post-merger emission represented as a Gaussian spectrum shown in Figure \ref{FigdedfBNS}. The BBH curves are scaled up by a factor of 4 to separate the three groups of curves, which is also the case in Figures \ref{OmegaSFR}-\ref{OmegaSimBHmass}.}
\label{FigOmegaCBCana}
\end{figure}

Figure \ref{FigOmegaCBCana} compares three models of $\Omega_{\rm{GW}}(f)$, using a NS (BH) mass of $1.33 \hspace{0.5mm} (7.8) \hspace{0.5mm} M_{\odot}$ and zero BH spin, and assuming that sources of each population have the same mass/spin values:\newline
1) A semi-analytical model calculated using equation (\ref{omeg3}) with complete waveforms described in subsection \ref{Newwaveforms};\newline
2) A Newtonian model based on equations (\ref{omeg3}) and (\ref{dEdf}), and assuming $f_{\rm{r}}^{\rm{max}}= f_{\rm{LSO}}$;\newline
3) A simple power law model based on equations (\ref{J23HB06}) and (\ref{omegCBC0}) with an upper frequency cutoff $f_{\rm{LSO}}/5$. Note that: a) an exact power law relation applies only for $f \leqslant f_{\rm{LSO}}/(1+z_{\ast})$ with $z_{\ast}=6$; we empirically set the cutoff at $f_{\rm{LSO}}/5$ since the function inside of the integral in equation (\ref{J23CBC}) has negligible values for $z\geqslant 4$; b) As mass distributions are not considered here, $\langle M_c^{5/3} \rangle$ in equation (\ref{omegCBC0}) becomes $M_c^{5/3}$ with $M_c$ determined by two component masses mentioned above.

The following features can be observed from Figure \ref{FigOmegaCBCana}: a) Newtonian models can be perfectly described by simple power law models up to $f_{\rm{LSO}}/5$, about 300 Hz, 80 Hz and 60 Hz for BNS, BH-NS and BBH respectively, as suggested in the previous paragraph; b) Semi-analytical models start to drop slightly below a $f^{2/3}$ power law from a few tens Hz due to PN amplitude corrections; c) Newtonian models give incorrect peaks and the followed abrupt decline because of the exclusion of post-inspiral emission.

For BNS, we specifically show that: a) the power law index of $\Omega_{\rm{GW}}(f)$ drops from $2/3$ ($f \lesssim 100$ Hz) to $1/3$ before peaking at around 1-2 kHz; b) if the post-merger emission is included in the form of Gaussian spectra, the peak of $\Omega_{\rm{GW}}(f)$ can be considerably enhanced while the low frequency part ($\lesssim 300$ Hz) stays at the same level. We will show, however, in section \ref{CalSNR} that the contribution from post-merger emission to the background is unlikely to be detectable even with ET.

\begin{figure}
\includegraphics[width=0.48\textwidth]{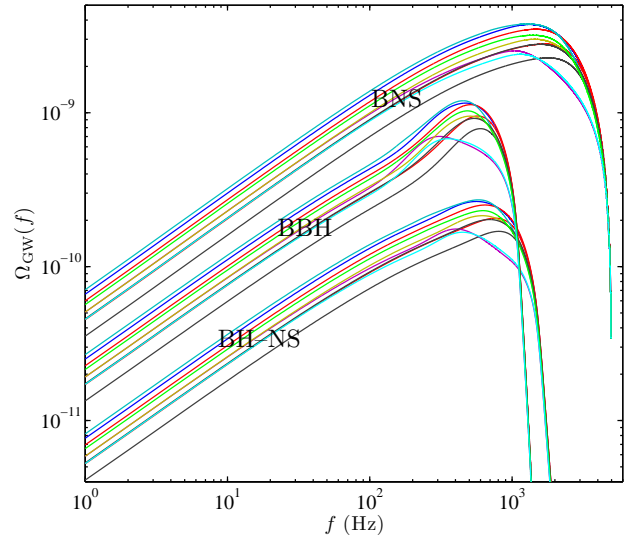}
\caption{As in Figure \ref{FigOmegaCBCana}, but only shows the semi-analytical models calculated for 10 different rate evolution models based on the five CSFRs shown in Figure \ref{ezSFR} and two minimum delay times $t_{\rm{min}} = 10, 100$ Myr. There is a degeneracy between CSFR model and $t_{\rm{min}}$ below 100 Hz and the unique signature at around the peaks is due to the CSFR model in \citet{SFR_GRB11}.}
\label{OmegaSFR}
\end{figure}

Figure \ref{OmegaSFR} shows $\Omega_{\rm{GW}}(f)$ of the semi-analytical models using 10 different forms of $e(z)$ based on the five CSFR models and two minimum delay times $t_{\rm{min}} = 10, 100$ Myr. Two main results are: a) for each population, different curves follow the same gradient up to 100-200 Hz. This is in agreement with our derivation in subsection \ref{smodelCBC}, where we have shown within this frequency range effects of the CSFR and delay times are linear. There is a degeneracy between the CSFR and $t_{\rm{min}}$: to break this degeneracy a fully reconstructed $e(z)$ and precise CSFR measurements are required; the former could become possible if we can efficiently detect most of the individual events out to high redshift, e.g., as we will show in section \ref{Subinds} for the BBH population; b) the only distinguishable feature comes from the adoption of CSFR model of \citet{SFR_GRB11}; the relatively high CSFR from $z=3$ up to $z=15$ shifts the peaks of $\Omega_{\rm{GW}}(f)$ to lower frequencies and suppresses the post-peak amplitudes.

\section{Monte-Carlo simulation}
\label{Sim}
In this section we describe a Monte-Carlo simulation approach to calculate $\Omega_{\rm{GW}}(f)$ of a general AGWB. This will allow us to investigate two important aspects of the CBC background in the next two sections:\newline
1) The dependency of $\Omega_{\rm{GW}}(f)$ on NS/BH mass distributions;\newline
2) How much of the CBC background can be removed through single-source detections to allow greater accessibility to primordial GWBs from the Big Bang.

Combining equations (\ref{omeg2})-(\ref{dVdz}) and (\ref{flux2}) yields:
\begin{equation}
\Omega_{\rm{GW}}(f) = \frac{1}{\rho_{c}} \frac{r_{0}}{H_{0}} \frac{2\pi^2 c^3}{G} \int_{z_{\rm{min}}}^{z_{\rm{max}}} f^3 \langle|\tilde{h}(f)|\rangle^2 g(z)\, {\rm{d}}z \label{Omeghf},
\end{equation}
where we have defined
\begin{equation}
g(z) = \frac{r_{z}^2 e(z)}{(1+z)\sqrt{\Omega_{\Lambda}+\Omega_{m}(1+z)^{3}}} \, \label{gfunz},
\end{equation}
assuming that there is no correlation between the rate evolution and source intrinsic parameters.

The discrete version of the integration in equation (\ref{Omeghf}) is a sum over events distributed in redshift, leading to:
\begin{eqnarray}
\Omega_{\rm{GW}}(f)\hspace{-1mm} &=& \hspace{-1mm} \frac{1}{\rho_{c}} \frac{r_{0}}{H_{0}} \frac{2\pi^2 c^3}{G} \times \label{Omegdisc}\\&& \hspace{-2mm} \frac{1}{N_{\rm{mc}}} \sum_{i=1}^{N_{\rm{mc}}} f^3(\Theta_i, z_i) \langle|\tilde{h}(f; d_L^i, \Theta_i)|\rangle^2 \frac{g(z_i)}{P(z_i)}\nonumber,
\end{eqnarray}
where $i$ denotes the $i$-th event; $\Theta_i$ contains the intrinsic source parameters, which in our case includes binary component masses $m_1^i$ and $m_2^i$, and BH spin parameters $\chi_1^i$ and $\chi_2^i$; the parameter $d_L$ is the luminosity distance, given by $r_{z} (1+z)$; the function $P(z)$ is the probability distribution function of source redshift $z$; and $N_{\rm{mc}}$ is the number of events in our Monte-Carlo simulation, chosen to be $10^6$ -- the approximate expected number of BNS merger events within $z_{\ast}$ in one-year observation. Note that: a) $\Omega_{\rm{GW}}(f)$ does not depend on $N_{\rm{mc}}$ or an observation time; b) $1/[N_{\rm{mc}} P(z)]$ plays the role of ${\rm{d}}z$ in the integration of equation (\ref{Omeghf}) -- $1/P(z)$ is essentially a weight used when calculating the average over individual sources; it simply becomes the length of integration $(z_{\rm{max}} - z_{\rm{min}})$ without any prior knowledge of source redshift distribution, e.g., for semi-analytical integrations of section \ref{CBCsemi} and other similar studies.

To investigate the detectability of individual events and show how much of the CBC background can be removed through the subtraction of detected events (in section \ref{Subinds}), we adopt the so-called \textit{effective} distance $D_{\rm{eff}}$, which is related to $d_L$ through \citep{AllenFindChirp05}:
\begin{equation}
D_{\rm{eff}} = d_L \left [ F_{+}^2 \left(\frac{1+\cos ^{2}\iota}{2}\right)^{2} + F_{\times}^2 \cos ^{2}\iota \right]^{-1/2},
\label{Deff}
\end{equation}
where $F_{+}$ and $F_{\times}$ are the antenna pattern functions for $+$ and $\times$ polarized GWs respectively, depending on source position with respect to the detector (described by the right ascension $\theta$ and declination $\phi$ of the source) and the polarization angle $\psi$; $\iota$ is the inclination angle. When averaging over uniformly distributed $\theta$, $\phi$ and $\psi$, one obtains $\langle F_{+}^2\rangle = \langle F_{\times}^2\rangle = \sin^{2} \zeta/5$ with $\zeta$ being the opening angle between the two arms of the laser interferometer \citep[see, e.g.,][]{magg00}.

An initial step in performing a Monte-Carlo simulation is to construct probability distribution functions of parameters. Here we use the NS/BH mass and BH spin distributions given in Table \ref{tb1} and further assume that $m_1$ and $m_2$, $\chi_1$ and $\chi_2$ are uncorrelated. The parameters $\cos \theta$, $\phi/\pi$, $\psi/\pi$ and $\cos \iota$ are all uncorrelated and uniformly distributed over $[-1, 1]$, where the consideration with $\cos \theta$ and $\cos \iota$ ensures that individual sources and the direction of their orbital angular momentum are uniformly distributed on a spherical surface. The function $P(z)$ is obtained by normalizing the differential event rate given in equation (\ref{dR}).

Our final results are obtained using an average of 10 independent realisations of the Monte-Carlo simulation. Numerical error in our simulation, defined as the relative variation (between each realisation and the average) of reference values of $\Omega_{\rm{GW}}(f)$ at 100 Hz, are within a few percent. As individual sources contribute to the background through a $f^{2/3}$ power law below 100 Hz, the outputs of different realisations vary by a small linear factor in magnitudes of $\Omega_{\rm{GW}}(f)$. Note that our results represent the \textit{average} background energy spectra. The actual background signal can deviate considerably from the \textit{average} depending on the time-frequency properties as we will discuss in section \ref{resoveCBC}.

We shall present our results as both: a) a full background -- calculated using equation (\ref{Omegdisc}) for each population, which is the background signal we will be searching for; b) a residual foreground -- in the summation of equation (\ref{Omegdisc}) individual events above a given detection threshold (see section \ref{Subinds}) are discarded, which represents the residual noise due to sub-threshold sources. Result a) will be presented in section \ref{NumResults}, and a) and b) are compared in section \ref{Subinds}. For completeness, we also present an example of a simulated time series due to the BNS population in section \ref{resoveCBC}.

\subsection{Backgrounds encoded with mass distributions}
\label{NumResults}
We compare the numerical results of $\Omega_{\rm{GW}}(f)$ for each CBC population, with the semi-analytical model presented in subsection \ref{SemiModels} and a simple power law model described in subsection \ref{smodelCBC}.

\begin{figure}
\includegraphics[width=0.48\textwidth]{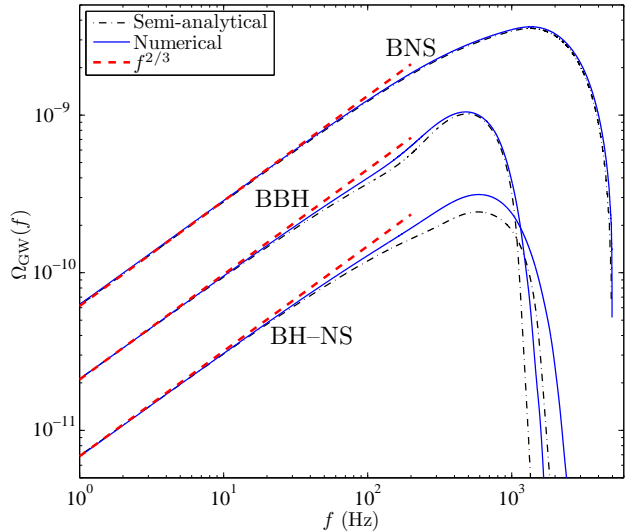}
\caption{The CBC background energy spectra calculated numerically assuming Gaussian mass distributions are compared with the semi-analytical models and simple power law models $f^{2/3}$ (up to 200 Hz) shown in Figure \ref{FigOmegaCBCana}.}
\label{OmegaSimAna}\vspace{-3mm}
\end{figure}

\begin{figure}
\includegraphics[width=0.48\textwidth]{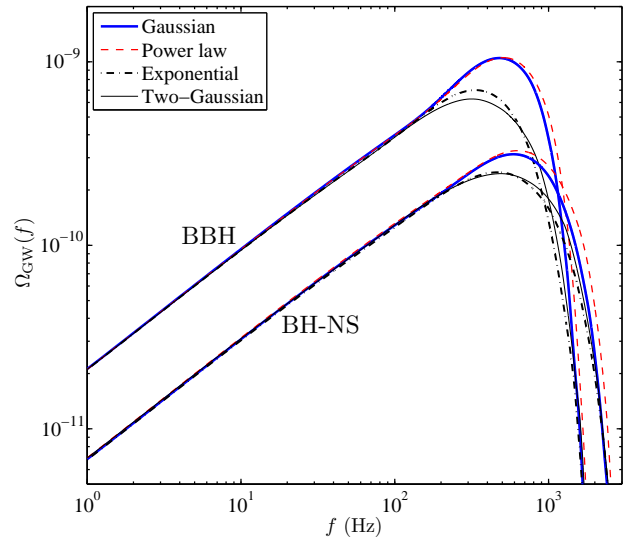}
\caption{As in Figure \ref{OmegaSimAna}, but showing only the numerical results of BH-NS and BBH for 4 models of BH mass distribution. Curves are scaled linearly to the same $\langle M_c^{5/3} \rangle$ as that of a Gaussian distribution (see Table \ref{tb2}).}
\label{OmegaSimBHmass}\vspace{-2mm}
\end{figure}

Figure \ref{OmegaSimAna} shows such a comparison in the case of Gaussian mass distributions. The very narrow distribution of NS masses has negligible influence on the BNS background -- the numerical result perfectly matches the semi-analytical model except a slightly broader shape around the peak, while effects of a Gaussian BH mass distribution are moderately noticeable for BH-NS and BBH -- the reduction from the $f^{2/3}$ curve is partly alleviated due to the contribution from the merger-ringdown emission of more massive systems. Unless otherwise stated we adopt the numerical models assuming Gaussian mass distributions in the following sections.

Figure \ref{OmegaSimBHmass} illustrates the numerical models of BH-NS and BBH for different BH mass distributions. Curves are scaled according to the individual values of $\langle M_c^{5/3} \rangle$ as given in Table \ref{tb2} for each distribution to ensure that all have the same value as that of a Gaussian distribution. Two groups are clearly distinguishable above 200 Hz -- one containing low mass BHs only (Gaussian and power law) and another for a broader distribution when high mass BHs are also included -- different peak frequencies are due to variations of average total masses and slightly distinct spectral width comes from the different degree of concentration of the distribution. The amplitudes of $\Omega_{\rm{GW}}(f)$ up to 100 Hz are very similar to each other (within numerical errors), agreeing with the expected dependency on $\langle M_c^{5/3} \rangle$.

Two main conclusions from Figures \ref{OmegaSimAna} and \ref{OmegaSimBHmass} are: a) mass distribution plays a role only through $\langle M_c^{5/3} \rangle$ in the low-frequency ($\lesssim$ 100 Hz) power law part of the background energy spectrum. This was mentioned in \citet{WuCBC12}, and also confirmed independently in \citet{Kowalska12} where a power law relation was obtained using mass distributions derived from population-synthesis simulations; b) it could become possible to probe mass distributions through stochastic background measurements, e.g., peaks shown in Figure \ref{OmegaSimBHmass} (once measured) will provide information about the average total mass and the degree of concentration of the distribution.

In our calculations we do not consider more sophisticated distributions of BH spin and mass ratio. These two parameters play a minor role (compared with $M_c$) above a few tens Hz in our adopted waveforms. Therefore, our results will not be affected significantly as long as their true distributions are not highly asymmetrical. An additional effect due to orbital eccentricity is not relevant as the orbits of coalescing compact objects are expected to circularize before their GW signals enter the ground-based frequency window \citep{BrownEccen10}. We note, however, that dynamically formed BBHs, of which the population is not considered in the current work, may be highly eccentric and could merge before their orbits are circularized \citep{DynaBBH11}. As mentioned in \citet{BBH_Zhu11}, such a population, possibly with much higher average masses, could provide considerable contribution to a GWB.

\section{Issues on the detection}
\label{CalSNR}
In this section we first update previous estimates on the detectability of the CBC background for second and third generation terrestrial detectors, using improved background models. By considering practical issues in detection and parameter estimation, we further discuss the choice of $\Omega_{\rm{GW}}(f)$ templates for data analysis.

Following \citet{BBH_Zhu11}, we consider five advanced detectors -- advanced LIGO \cite[aLIGO;][]{aLIGO} at Hanford (H) and Livingston (L), advanced Virgo\footnote{http://wwwcascina.virgo.infn.it/advirgo/} (V) in Italy \citep{adVirgo}, KAGRA \citep[K, previously known as LCGT,][]{KAGRA} in Japan, and AIGO in Australia \citep{AIGO}, as well as ET\footnote{http://www.et-gw.eu/etsensitivities} for which we consider two configurations -- ET-B \citep{et-b} and ET-D \citep{et-d}. Note that the inclusion of IndIGO\footnote{http://www.gw-indigo.org} in India should have similar contribution as AIGO. Unless otherwise stated, we use the sensitivity curves of aLIGO for the zero detuning, high laser power configuration (see the public LIGO document T0900288 for details\footnote{ https://dcc.ligo.org/cgi-bin/DocDB/ShowDocument?docid=2974}), and of KAGRA for the broadband configuration\footnote{http://gwcenter.icrr.u-tokyo.ac.jp/en/researcher/parameter}. We assume that AIGO has the same sensitivity as aLIGO. The target sensitivities of these detectors are shown later in Figure \ref{ShSnComp}.

\subsection{Signal-to-noise ratios}
\label{SNRresults}
The optimum detection strategy for a stochastic GWB is to cross-correlate the outputs of two or more detectors \citep[see, e.g.,][]{AllenGWB99,magg00}. Strictly speaking, the CBC background is not a stochastic background in the sense that individual signals do not sufficiently overlap in time-frequency space, as suggested in \citet{Rosado11} and we will also discuss in section \ref{resoveCBC}. Nevertheless it has been shown, both theoretically \citep{drasco03} and experimentally \citep[through data analysis exprement on simulated data;][]{Tania_ETMDC}, that the cross correlation method works nearly optimally in the non-Gaussian regime, because through long time integration it is always possible to obtain a sufficiently large number of signals in a frequency interval and ``form'' a Gaussian background for which the cross correlation statistic applies. Therefore, we consider this standard method to assess the detectability of the CBC background for future detectors.

The optimal signal-to-noise ratio ($\mathrm{S/N}$) obtainable by two-detector cross correlation is given by \citep[e.g., equation 3.75 in][]{AllenGWB99}:
\begin{equation}
{(\mathrm{S/N})^2} =  2 T \int_0^\infty \> {\gamma^2 (f) S_{h}^2(f) \over S_{n1}(f) S_{n2}(f)} {\rm{d}}f\
\label{SNR},
\end{equation}
where $\gamma (f)$ is the normalized overlap reduction function, which accounts for the sensitivity loss due to the separation and relative orientation of the two detectors \citep{Flanagangamma93}. For co-located and co-aligned detectors, $\gamma (f) =1$. The one-sided noise power spectral densities of the two detectors are given by $S_{n1}(f)$ and $S_{n2}(f)$, and $T$ is the integration time (set to be one year). Note that we have substituted $\Omega_{\rm{GW}}(f)$ with the spectral density $S_h(f)$ through equation (\ref{Sh}) to obtain a more intuitive format. We use the $\gamma (f)$ for the 10 pairs of advanced detectors presented in \citet{Nishizawa09orf} and adopt the form of ET for two V-shaped detectors separated by $120^{\circ}$ \citep[see Figure 8 of][]{Tania_ETMDC}.

As we will be observing a GWB due to all possible contributions of CBC sources, we calculate the $\mathrm{S/N}$ of the total background from the three CBC populations without considering other types of sources. The background spectrum of the total CBC background is simply the sum of that of each population, as shown later in Figure \ref{OmegaCBCsimSub}.

We present in Table \ref{tb3} values of $\mathrm{S/N}$ calculated for advanced detectors. We consider three cases:\newline
1) Cross correlation between pairs of advanced detectors using real $\gamma (f)$ and the individual sensitivities of each detector; we additionally assume all detectors have the same sensitivity as aLIGO to evaluate the effect of $\gamma (f)$;\newline
2) Assuming $\gamma (f) =1$ for pairs of detectors with the sensitivity of either aLIGO, or KAGRA or advanced Virgo;\newline
3) An optimal combination of cross correlation statistics for 10 pairs of advanced detectors, for which $(\mathrm{S/N})^2$ is simply the sum of those calculated in case 1) \citep[see, e.g., equation 5.46 in][]{AllenGWB99}.\newline
Note that case 3) is mathematically simple but requires 5 advanced detectors to be simultaneously online.

\begin{table}
\begin{center}
\caption{\label{tb3} $\mathrm{S/N}$ of the CBC background for advanced detectors.}
\vspace{-2mm}
\begin{tabular}{lccccc}
\hline \hline
   Pair      & A-H    & A-K    & A-L    & A-V    & H-K \\
  \textit{a} & 0.43   & 0.11   & 0.46   & 0.10   & 0.06 \\
  \textit{b} & (0.43) & (0.17) & (0.46) & (0.18) & (0.10) \\
\hline
   Pair      & H-L    & H-V    & K-L    & K-V    & L-V \\
  \textit{a} & 1.05   & 0.12   & 0.02   & 0.12   & 0.14 \\
  \textit{b} & (1.05) & (0.23) & (0.03) & (0.29) & (0.25) \\
\hline
 Case & HH   & KK     & VV     & \textit{Ca} &\textit{Cb}   \\
      & 2.76 & 1.82   & 1.31   & 1.26         & 1.34   \\
\hline \hline
\end{tabular}
\end{center}
\vspace{-3mm}
\medskip{Notes: Assuming that the background is contributed by the three populations of CBC sources (BNS, BH-NS, and BBH), we present $\mathrm{S/N}$ of cross correlating different interferometer pairs from a worldwide network. Results are shown for: \textit{a} -- individual detector sensitivities (but note that A, H and L have the same aLIGO sensitivity); \textit{b} -- assuming all detectors have aLIGO sensitivity. The motivation for calculating \textit{b} is to investigate effects of the overlap reduction function $\gamma (f)$; HH, KK, and VV assume $\gamma (f)=1$ for aLIGO, KAGRA and advanced Virgo respectively; Values of \textit{Ca} and \textit{Cb} are just the square roots of the quadratic sum of the 10 values in \textit{a} and \textit{b} respectively, which can in principle be achieved by optimally combining measurements from multiple pairs of detectors. The improvement of \textit{Ca} and \textit{Cb} on H-L is not appreciable due to the suppressing effects of $\gamma (f)$.}
\end{table}

Our results show that: a) among the 10 pairs, H-L performs the best in terms of detecting a CBC background, giving a $\mathrm{S/N}$ of 1, while the lowest value of $\mathrm{S/N}$ is only 0.02. Assuming the aLIGO sensitivity for all detectors only increases the lowest value to 0.03; b) the improvement from combining the network of advanced detectors is only $\sim 30\%$ on the best performing pair H-L, while assuming $\gamma (f) =1$ for aLIGO increases the $\mathrm{S/N}$ by nearly 3 folds. This is well below the expectation that these two should give similar improvement \citep{WuCBC12,Kowalska12}. Such a pessimistic prospect is mainly due to effects of $\gamma (f)$. This has been pointed out in our previous studies \citep{r-mode2011,BBH_Zhu11} and will be discussed in more details below.

The property of $\gamma (f)$ is mainly described by its characteristic frequency $f_{\rm{char}}$, given by $f_{\rm{char}} = c/(2|\Delta X|)$ with $|\Delta X|$ the distance between two detectors, above which $\gamma (f)$ decays rapidly towards zero. Among the 10 pairs of detectors, H-L has the smallest separation ($|\Delta X| = 3000$ km), resulting in the highest $f_{\rm{char}}$ of 50 Hz, while values of $f_{\rm{char}}$ for other pairs vary from 10 Hz to 20 Hz \citep{Nishizawa09orf}. This, combined with the fact that advanced detectors have a low frequency seismic wall at about 10 Hz, can easily explain the very small $30\%$ improvement. Note that the overlap reduction function also depends on the relative orientation of the two detectors, and we refer interested readers to \citet{Nishizawa09orf} for discussion about the optimal configurations of (geographically separated) detector pairs.

For ET, the CBC background can be easily detected, with $\mathrm{S/N}$ of 178 (350), 19 (38), and 15 (30) assuming ET-B (ET-D) sensitivity for the BNS, BH-NS and BBH population respectively -- the factor of 2 increment from ET-D is due to greater sensitivity at frequencies below $\sim 20$ Hz. This implies that the detection prospects benefit significantly from improvement of low-frequency sensitivities (as also shown later in Figure \ref{MinOmegaLIGO}).

We note that the (optimistic) BNS post-merger contribution to the GWB, as shown in Figure \ref{FigOmegaCBCana}, results in a $\mathrm{S/N}$ of only 0.43 (0.46) for ET-B (ET-D), implying that detecting the imprint of BNS post-merger emission on a GWB requires a coalescence rate at least 5 times higher than the realistic value adopted in this study. On the other hand, we quantify PN effects with the difference in $\mathrm{S/N}$ between a simple power law model and the numerical model shown in Figure \ref{OmegaSimAna} (both are cutoff at 200 Hz). We find that PN amplitude corrections cause a reduction of $\mathrm{S/N}$ in the range of ($5.5\% - 8.5\%$) and ($2.1\% - 3.2\%$) for ET-B and ET-D respectively. ET-D is less sensitive to these effects as its best sensitivity is more concentrated at lower frequencies.

\subsection{Detection prospects for advanced detectors}
\label{DetAdvIFOs}
As the operation of advanced detectors is only two years away, it is now important to carefully assess the detection prospects of the CBC background, which represents one of the most (if not the most) promising background sources. We look at this issue in much more depth by considering variations in both source parameters and detector configurations.

For each population, $\Omega_{\rm{GW}}(f)$ scales linearly with $r_0$, for which the uncertainties are generally of orders of magnitude -- much larger than those of other parameters. To determine what possible combinations of $r_0$ the total CBC background will be accessible to advanced instruments, we simply scale the numerical models for different values of $r_0$, keeping all other parameters fixed. The considered range of $r_0$ (in $\rm{Mpc}^{-3} \hspace{0.5mm} \rm{Myr}^{-1}$) is $0.1 - 10$ (pessimistic to optimistic) for BNS, and $0.005 - 0.30$ (realistic to optimistic) for BBH, while the BH-NS rate is set to be the realistic 0.03 \citep[all values taken from Table 4 of][]{lsc_rate10}.

The motivation of our choice regarding BBH and BH-NS is two-fold: a) it was recently predicted, through population synthesis studies \citep{metal_BBHrate10,Dominik12StarTrack} and empirical estimation based on two observed BH-Wolf-Rayet star systems \citep{Bulik11_BBH}, that $r_0$ for BBH can plausibly be at the optimistic value adopted above; b) while the same population synthesis studies gave similar realistic rates of BH-NS \citep[see, e.g., Tables 2 and 3 in][]{Dominik12StarTrack}, a negligible coalescence rate for BH-NS was recently empirically determined by following the future evolution of Cyg X-1 \citep{BelczynskiBH-NS2011}. In the current analysis for advanced detectors, the contribution of the BH-NS population is nearly negligible at the chosen rate.

\begin{figure}
\includegraphics[width=0.48\textwidth]{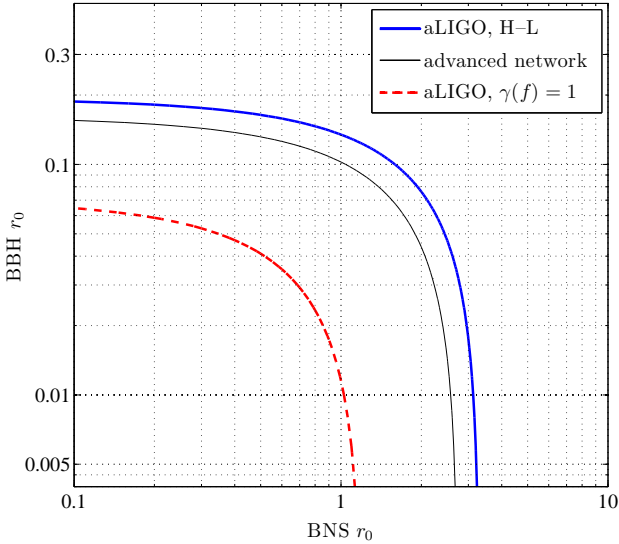}
\caption{The detectable ``rate space'' of a CBC background, assumed to be contributed by three populations (BNS, BH-NS and BBH), for ground-based advanced detectors. The local coalescence rate $r_0$ is in $\rm{Mpc}^{-3} \hspace{0.5mm} \rm{Myr}^{-1}$ with $r_0$ for BH-NS fixed at 0.03 (see text). A signal-to-noise ratio greater than 3 for one year observation can be obtained for rates above each curve assuming: a) cross correlation with aLIGO H-L; b) an ideal case of $\gamma (f) =1$ for two aLIGO detectors; c) an optimal combination of measurements of 10 pairs advanced detectors (advanced network). Note that this background will be easily detectable for ET, i.e., the corresponding curve is well below the origin of the two axes.}
\label{SNR3r0}
\end{figure}

Figure \ref{SNR3r0} shows the detectable ``rate space'' for advanced detectors: a $\mathrm{S/N}$ threshold of 3 is used to indicate detections, which corresponds to $95\%$ detection rate and $5\%$ false alarm rate. Note that: a) we have taken the integer 3 for convenience, while the accurate number is 3.29, given by $\sqrt{2} [\rm{erfc}^{-1}(2\beta)-\rm{erfc}^{-1}(2\varsigma)]$ assuming a false alarm rate $\beta = 5\%$ and a detection rate $\varsigma = 95\%$. Here $\beta$ and $\varsigma$ need not sum to 1 \citep[see][for details]{AllenGWB99}; b) one can lower the threshold on $\mathrm{S/N}$ with the cost of increasing $\beta$, or decreasing $\varsigma$ or both. For example, a threshold of 2.56 may be used if one chooses $\beta = 10\%$ and $\varsigma = 90\%$.

In the case of $\gamma (f) =1$ for aLIGO, our results are consistent with \citet{WuCBC12}, implying that the BNS population alone may produce a detectable background signal at the realistic coalescence rate. However, an important point here is that for a worldwide network of advanced detectors, the requirement for a detection is more than twice stronger. This motivates us to consider all three populations as a whole as they will be observed in reality. Figure \ref{SNR3r0} shows that some combinations of the BNS population and the BBH population can form a detectable GWB, while both of their individual contributions alone are not sufficient for detection. In practice, if a $\Omega_{\rm{GW}}(f) \sim f^{2/3}$ power law background has been detected, one would certainly wish to determine the relative contribution from every possible population.

The possible variation in $\langle M_c^{5/3} \rangle$ (for BBH and BH-NS systems) and the effects of CSFR and delay times (which can be represented using the parameter $J_{2/3}$) are not considered in Figure \ref{SNR3r0}. Combining the simple power law model given by equation (\ref{omegCBC0}), which will be shown to be a good approximation in the next subsection, with equations (\ref{Sh}) and (\ref{SNR}) we have a simple relation:
\begin{equation}
{\mathrm{S/N}} =  C_{2/3} \sum_{k=1}^{3} r_0^{k} \langle M_c^{5/3} \rangle ^{k} J_{2/3}^{k}\
\label{SNRscaling},
\end{equation}
where the indices $k=1, 2, 3$ denote the three CBC populations and $C_{2/3}$ is a constant depending only on detector sensitivity and $\gamma (f)$ for different detector pairs used in cross correlation. For convenience we have omitted the division by the corresponding reference values as in equation (\ref{omegCBC0}) for the three parameters. Note that one needs to set the upper frequency limit of the integration in equation (\ref{SNR}) at 100 Hz so that the above equation is representative of results obtained using numerical models of $\Omega_{\rm{GW}}(f)$. In fact, our Figure \ref{SNR3r0} can be easily reproduced by using equation (\ref{SNRscaling}) together with values of $\langle M_c^{5/3} \rangle$ and $t_{\rm{min}}$ (to obtain the parameter $J_{2/3}$ through equation (\ref{J23HB06})) given in Table \ref{tb2}.

Varying the values of $\langle M_c^{5/3} \rangle$ for BBH and BH-NS from those of a Gaussian distribution (as assumed in Figure \ref{SNR3r0}) to the highest entries in Table \ref{tb2} increases the total $\mathrm{S/N}$ by $5\%$ ($40\%$) assuming a BBH coalescence rate of $r_0 = 0.005 \hspace{0.5mm} (0.3) \hspace{0.5mm} \rm{Mpc}^{-3} \rm{Myr}^{-1}$ and realistic values of $r_0$ for both BNS and BH-NS. Such a increment of $\mathrm{S/N}$ is smaller than $40\%$ for higher coalescence rates of BNS and BH-NS. Therefore, our Figure \ref{SNR3r0} does not change appreciably for variations of $\langle M_c^{5/3} \rangle$ given the current observational BH mass estimates. Meanwhile, effects of CSFR and delay times could moderately degrade (i.e., no more than a factor of 2) the detection prospects, as our current choice gives $J_{2/3} = 2.3$ for the dominant BNS population, which is close to the high end of the range $(1.3 - 2.6)$ obtained in section \ref{smodelCBC}.

Note that for a putative population of dynamically formed BBHs in dense stellar clusters \citep[see, e.g.,][]{Sadowski08} or for the same field population (as considered in this study) but assumed to be formed in low metallicity environments \citep{Dominik12StarTrack}, a much larger average chirp mass $\langle M_c \rangle$ up to about $\sim 20 M_{\odot}$ (in comparison to $\langle M_c \rangle \simeq 7 M_{\odot}$ for the mass distribution used in this work) was suggested to be possible\footnote{In the study of \citet{Dominik12StarTrack}, the authors found that a larger average chirp mass is associated with a higher coalescence rate for BBH in low metallicity environments, resulting in BBHs dominating the whole CBC population.}. To allow for these possibilities, a plot of detectable $r_{0} - \langle M_{c} \rangle$ space is useful. Such illustrative studies, which only apply to a single population, have been presented for the BBH population in \citet{BBH_Zhu11}, and for each of the three CBC populations in \citet{WuCBC12} -- for a given $\mathrm{S/N}$ threshold, the scaling relation $r_0 \sim  \langle M_c \rangle^{-5/3}$ was shown to be a good approximation\footnote{Curves for advanced detectors in Figure 5 of \citet{BBH_Zhu11} underestimate the detectability by a factor of 4 due to the use of an old version of aLIGO sensitivity and one additional factor of $(1+z)$ in the calculation of $\Omega_{\rm{GW}}(f)$.}. At a coalescence rate of the order $10^{-3} \hspace{0.5mm} \rm{Mpc}^{-3} \rm{Myr}^{-1}$ \citep[see, e.g.,][]{Cmiller09}, dynamical formation scenarios should have similar contributions to a GWB as the field population of BBHs considered in this study (at the realistic rate) and thus will not improve considerably the detection prospects for advanced detectors. Such a back-of-the-envelope argument also applies to binaries involved with one or two intermediate mass BHs - much lower rates cancel out the advantage of higher masses \citep[see, e.g., Tables 8-10 of][]{lsc_rate10}. Despite of the involved uncertainties, these systems should be a more interesting source for single event searches/detections (at least for advanced detectors), from which their very existence will be tested or the associated coalescence rates can be stringently constrained.

Looking forward to the advanced detector era, it is now crucial to investigate how the detection prospects of the CBC background (which could be the first to be detected) can be enhanced. For co-located detectors, techniques to remove correlated environmental and/or instrumental noises will be required, and are currently being developed in the LIGO/Virgo collaboration \citep{LIGOHH08}. Provided that no co-located instruments are available, detection of a GWB from CBC events will require higher coalescence rates than what are presently thought to be realistic, i.e., $r_0 = 3 \hspace{0.5mm} (0.2) \hspace{0.5mm} \rm{Mpc}^{-3} \hspace{0.5mm} \rm{Myr}^{-1}$ for BNS (BBH), or alternatively given the realistic rates an integration time of 4 years to obtain a $\mathrm{S/N}$ of 2, which was assumed as a threshold in \citet{WuCBC12}. We note that a single-detector auto-correlation approach was recently proposed to be comparable in $\mathrm{S/N}$ to what is achievable by cross correlation of two co-located and co-aligned detectors \citep{Tinto12}, of which the feasibility needs to be further tested in realistic data analysis experiments.

The above results have assumed standard versions of design sensitivities for advanced detectors. In practice, detectors can be tuned to different configurations for various purposes, e.g., to allow optimization for different searches. As the aLIGO H-L pair gives the majority of contribution to the network $\mathrm{S/N}$ for a CBC background, we consider here four additional tuning options of aLIGO (data for the corresponding sensitivities are available publicly at the link given in the beginning of this section, and we refer interested readers to the LIGO document T0900288 therein for descriptions and technical details): a) Zero-detuning, low laser power; b) Optimal NS-NS, which is optimized to the BNS inspiral search; c) Optimal BH-BH, which is optimized for 30-30 solar mass BBH inspirals; d) High frequency, which has a narrowband tuning at 1 kHz.

We re-calculate the $\mathrm{S/N}$ of the total CBC background for H-L using the additional four sensitivity curves, and obtain 1.18, 0.83, 1,49 and 1.23 for a), b), c) and d) respectively, in comparison to 1.05 for the standard configuration (zero-detuning with high laser power). This shows that modest improvement of low-frequency sensitivity provide considerable enhancement in $\mathrm{S/N}$, which is comparable to or even greater than that due to the combination of multiple detector pairs (again for the currently proposed network). The largest value of $\mathrm{S/N}$ (for one year observation), which comes from the adoption of c), implies that $\mathrm{S/N} = 3$ is achievable with an integration time of 4 years in comparison to 8 years for the standard option.

\begin{figure}
\includegraphics[width=0.48\textwidth]{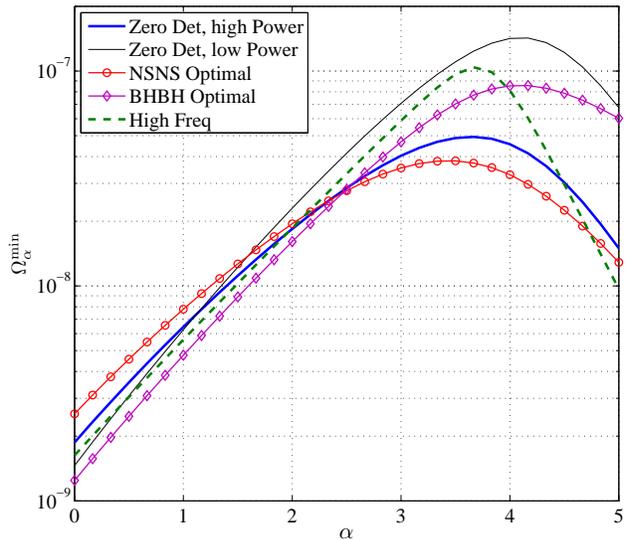}
\caption{The minimum detectable energy density $\Omega_{\alpha}^{\rm{min}}$ for generic power law GWB models (with indices $\alpha$) -- $\Omega_{\rm{GW}}(f) = \Omega_{\alpha} (f/100\hspace{0.5mm} \rm{Hz})^{\alpha}$, for one year observation using the aLIGO H-L pair. We assume a $\mathrm{S/N}$ threshold of 3, and consider five tuning options for aLIGO: zero-detuning with high/low laser power, optimized for searches of NS-NS/BH-BH inspirals and one with high frequency narrowband tuning. We refer interested readers to the public LIGO document T0900288 for descriptions and technical details about aLIGO tunings.}
\label{MinOmegaLIGO}
\end{figure}

Based on the above analysis, we suggest that the optimal BH-BH option offers an appreciable increase in sensitivities of stochastic searches. To make this suggestion more accessible to experimentalists and to extend the above comparison to generic power law GWB models -- $\Omega_{\rm{GW}}(f) = \Omega_{\alpha} (f/100\hspace{0.5mm} \rm{Hz})^{\alpha}$, we show in Figure \ref{MinOmegaLIGO} the minimum detectable energy density $\Omega_{\alpha}^{\rm{min}}$ for aLIGO H-L considering different tuning options. The values of $\Omega_{\alpha}^{\rm{min}}$ can be easily obtained by setting a threshold on $\mathrm{S/N}$ and solving the equality given by equation (\ref{SNR}). We take the integration range from 10 Hz to 1 kHz, and consider the range (0 -- 5) for $\alpha$. The curves in Figure \ref{MinOmegaLIGO} represent the upper limits obtainable by aLIGO, which apply to primordial GWBs (in addition to AGWBs), e.g., $\alpha = 0$ in many early-Universe scenarios \citep[see, e.g., Figure 2 in][]{LIGO-SGWB-limit}.

\subsection{The construction of $\Omega_{\rm{GW}}(f)$ templates}
\label{OmegaTempl}
In the previous sections, we have shown that:\newline
1) for $f \lesssim 100 $ Hz, the power law model given by equation (\ref{omegCBC0}) is a good approximation and requires only three parameters. The power law relation holds for three populations and thus for the total background as well;\newline
2) above 100 Hz, PN corrections become more notable, and different behaviors are expected from other effects such as CSFR and mass distributions, making it difficult to predict the background spectral properties.\newline
In this subsection we show that the power law model is sufficient to be used as search templates for a CBC background and is also useful for parameter estimation of the coalescence rate and average chirp mass (information other than these two quantities can only be extracted from measurements of high-frequency peaks).

\begin{figure}
\includegraphics[width=0.48\textwidth]{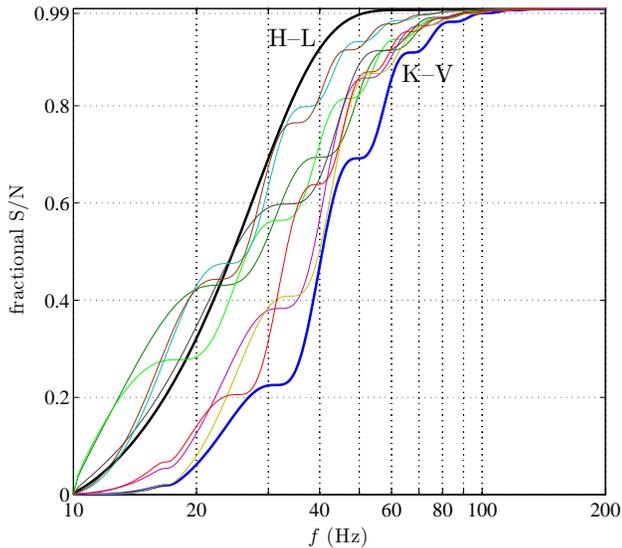}
\caption{The fractional signal-to-noise ratio ($\mathrm{S/N}$) as a function of upper frequency limits for the 10 pairs of advanced detectors for the BNS background. The ``fastest'' and ``slowest'' to accumulate $99\%$ of the total $\mathrm{S/N}$ are highlighted with thick lines, corresponding to cross-correlating aLIGO H-L and KAGRA-advanced Virgo (K-V) respectively.}
\label{fracSNRBNS}
\end{figure}

Figure \ref{fracSNRBNS} illustrates the fractional $\mathrm{S/N}$ as a function of upper frequency limits for the 10 pairs of advanced detectors for the BNS background. We see the $\mathrm{S/N}$ has saturated below 100 Hz due to the suppressing effects of $\gamma (f)$ and the fact that the background is ``red'', i.e., $S_{h}(f) \sim f^{-7/3}$. Quantitatively, a fraction of $99\%$ of the total $\mathrm{S/N}$ can be achieved up to 51 Hz and (at most) 98 Hz by cross-correlating H-L and K-V respectively. Such upper frequencies are slightly higher for ET-B (133 Hz) or assuming $\gamma (f) =1$ for aLIGO (128 Hz), and could be even lower for ET-D (47 Hz), as also noted in \citet{Kowalska12}. The exactly same values are obtained in the cases of BH-NS and BBH due to the similarity in $\Omega_{\rm{GW}}(f)$ below 200 Hz.

We then quantify the effectiveness of the proxy of a power law model to the CBC background below 100 Hz by looking at the following quantity:
\begin{equation}
\langle {\rm{S}} \rangle = \frac{T}{2} \, \lambda \int_0^{100} \> {\gamma^2 (f) S_{h}(f) S'_{h}(f) \over S_{n1}(f) S_{n2}(f)} {\rm{d}}f\
\label{Saveg},
\end{equation}
which gives the mean value of the cross-correlated signal, with $\lambda$ the normalization constant to ensure $\langle {\rm{S}} \rangle = \Omega_{\alpha} T$ for a power GWB with $\Omega_{\rm{GW}}(f) = \Omega_{\alpha} f^{\alpha}$ \citep{AllenGWB99}. Here $S_{h}(f)$ is the ``true'' spectral density of the background, assumed to be that given by our numerical models; $S'_{h}(f)$ corresponds to the template adopted in stochastic background searches. A simple power law template results in an overestimation of $\langle {\rm{S}} \rangle$ within $2\% - 5\%$ for the three CBC populations for 10 pairs of advanced detectors. This can be further reduced by up to $1\%$ by decreasing the upper cutoff frequency from 100 Hz to 50 Hz.

Overall, we suggest that a simple power law model for the CBC background as given by equation (\ref{omegCBC0}) with an appropriate upper frequency cutoff at 50-100 Hz is sufficient for detection and the followed-by parameter estimation of average masses and coalescence rates using ground-based interferometers. For third-generation detectors like ET, however, more accurate models, such as those presented in subsections \ref{SemiModels} and \ref{NumResults}, will be required to extract information such as CSFR, PN effects, and mass distributions.

\section{A foreground formed by sub-threshold CBC events}
\label{Subinds}
When searching for primordial GWBs from the early Universe, AGWBs formed by more recent sources could act as contaminating foregrounds. One resolution to this problem is to subtract individually detected signals from the data. This has been demonstrated for the proposed Big Bang Observer, which has a sufficiently good sensitivity that it can resolve and thus subtract away almost all BNS inspirals in the Universe from the overall background \citep{BBOsubtCutler06}. We refer interested readers to \citet{BBOsubtCutler06} for details of the method and related practical issues, and we simple apply this method to ET to estimate the ``residual'' foreground from sub-threshold CBC events\footnote{As the detection horizon of advanced detectors such as aLIGO is at most $z \sim 0.4$ for BBH systems \citep{lsc_rate10}, a subtraction process can not reduce the foreground appreciably.}.

The optimal method to detect signals with known waveforms is through matched filtering, for which the optimal (single-event) signal-to-noise ratio, $\rho$, is given by \citep[see, e.g.,][]{lsc_rate10}:
\begin{equation}
\rho^2 = 4 \,\int_0^{f_{\rm{max}}} \frac{|\tilde{h}(f)|^2}{S_{n}(f)} {\rm{d}}f\,
\label{SNRsingle},
\end{equation}
where $f_{\rm{max}}$ is the maximum observed frequency, depending on source redshift, component masses and spins (if applicable). We use the ET antenna pattern function, which goes to equation (\ref{Deff}) for $D_{\rm{eff}}$ and determines the overall amplitudes of $|\tilde{h}(f)|$, for a triangle configuration including three V-shaped detectors \citep[see, e.g., equation (24) of][]{Tania_ETMDC}. Note that the spectral density of the CBC background is well below that of the instrumental noise of ET even at optimistic rate estimates. Therefore, we do not need to consider $S_{h}(f)$ as an additional contribution to $S_{n}(f)$ in equation (\ref{SNRsingle}), whereas one must do so in the case of the Big Bang Observer \citep{BBOsubtCutler06}.

\begin{table}
\begin{center}
\caption{\label{tb4} Detection rates ($N_{\rm{det}}$) of CBC sources for ET.}
\vspace{-2mm}
\begin{tabular}{lcccccc}
\hline \hline
                      & \multirow{2}{*}{CSFR} & $N_{\rm{tot}}$ &  \multicolumn{4}{|c|}{$N_{\rm{det}}$ (${\rm{yr}}^{-1}$)} \\
                      &      &   (${\rm{yr}}^{-1}$)   &  G   & P  & E  & TG \\
\hline
 \multirow{4}{*}{BNS} &  \multirow{2}{*}{HB} &  \multirow{2}{*}{255} & 30.4 & ... & ...  & ... \\
                      &                      &                      & (31.5) & ... & ...  & ... \\
                      &  \multirow{2}{*}{RE}  &  \multirow{2}{*}{335} & 18.5 & ... & ...  & ... \\
                      &                      &                      & (19.4) & ... & ...  & ... \\
\hline
 \multirow{4}{*}{BH-NS}& \multirow{2}{*}{HB}  & \multirow{2}{*}{7.14} & 3.20  & 3.10  & 3.45   & 3.42  \\
                      &                      &                      & (3.27) & (3.17) & (3.53) & (3.50) \\
                      & \multirow{2}{*}{RE}  &  \multirow{2}{*}{9.59} & 3.07 & 2.90  & 3.36   & 3.33  \\
                      &                      &                      & (3.27) & (2.99) & (3.45) & (3.42) \\
\hline
 \multirow{4}{*}{BBH} &  \multirow{2}{*}{HB}  & \multirow{2}{*}{1.07} & 0.95  & 0.93  & 0.979   & 0.97  \\
                       &                      &                      & (0.96) & (0.94) & (0.983) & (0.98) \\
                      &  \multirow{2}{*}{RE}  & \multirow{2}{*}{1.49} & 1.22  & 1.18  & 1.277   & 1.26  \\
                       &                      &                      & (1.23) & (1.19) & (1.284) & (1.27) \\
\hline \hline
\end{tabular}
\end{center}
\vspace{-3mm}
\medskip{Notes: All values are in $10^4$ and have assumed the realistic coalescence rates (see Table \ref{tb2}) adopted from \citet{lsc_rate10}. $N_{\rm{tot}}$ is the total event rate up to $z=6$ or $z=15$ for CSFR models of HB \citep{HB06} or RE \citep{SFR_GRB11} respectively. We scale the number of events above the detection threshold ($\rho \geqslant 8$) in the Monte-Carlo simulation (see section \ref{Sim}) according to $N_{\rm{tot}}$/$N_{\rm{mc}}$ to obtain the detection rate $N_{\rm{det}}$. We have considered NS/BH mass distributions as described in Table \ref{tb1} -- Gaussian (G), Power law (P), Exponential (E) and Two-Gaussian (TG); and adopted ET sensitivities of two configurations -- ET-B and ET-D (values are given in parentheses). ET-D gives slightly higher detection rates due to a greater low-frequency ($f \lesssim 20$ Hz) sensitivity.}
\end{table}

We calculate $\rho$ for each of the simulated CBC events in our Monte-Carlo simulation as described in section \ref{Sim}: those loudest events resulting in $\rho \geqslant \rho_{\rm{th}} = 8$ are discarded (termed with ``subtraction") to estimate a residual noise. Before moving forward to discussions of ET's potential in removing the CBC background through a subtraction process, we present in Table \ref{tb4} ET detection rates (which are conveniently obtained in our simulations) of CBC sources for completeness. The calculations improve the approximation (for advanced detectors) used in \citet{lsc_rate10} with the following considerations: a) cosmic evolution of coalescence rates, the standard $\Lambda$CDM cosmology and cosmological redshifts; b) observational NS/BH mass distributions; c) complete waveforms that include PN amplitude corrections. While these effects may not be important for detection rate predictions for advanced detectors as discussed in \citet{lsc_rate10}, they must be considered for ET due to its 1000 times larger accessible volume.

Based on results presented in Table \ref{tb4}, we find that the realistic CBC detection rates for ET are $10^5$ (BNS), $10^4$ (BH-NS) and $10^4$ (BBH) given the current realistic coalescence rate predictions \citep[see][for details]{lsc_rate10}. Note that ET will have an overall detection efficiency (defined as $N_{\rm{det}}/N_{\rm{tot}}$) of $\sim 10\%$ (BNS), $\sim 40\%$ (BH-NS) and $\sim 85\%$ (BBH), which is independent of $r_0$ and weakly dependent on coalescence rate evolution and sensitivity models as shown in Table \ref{tb4}.

\begin{figure}
\includegraphics[width=0.48\textwidth]{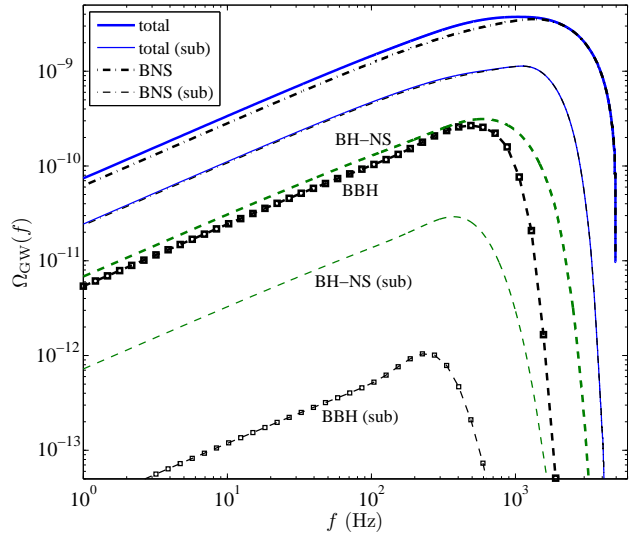}
\caption{The energy density parameter $\Omega_{\rm{GW}}(f)$ of the \textit{total} CBC background and its contributions by the \textit{BNS}, \textit{BH-NS}, and \textit{BBH} populations, without and with (labeled as \textit{sub}) the subtraction of individually detectable sources for ET assuming ET-B sensitivity (the results are essentially the same for ET-D sensitivity). Note that: a) the total residual ``noise" is dominated by sub-threshold BNS merger events; b) ET will be able to reduce the foreground level by a factor of about 2, 10 and 200 from the BNS, BH-NS and BBH population respectively; c)the reduction shown here is optimistic because of the assumption that every theoretically detectable single signals will be perfectly subtracted \citep[see][for detailed discussion about potential subtraction errors]{BBOsubtCutler06}.}
\label{OmegaCBCsimSub}
\end{figure}

Figure \ref{OmegaCBCsimSub} compares the results of $\Omega_{\rm{GW}}(f)$ calculated using equation (\ref{Omegdisc}) without and with the subtraction of individually detectable events. We see that ET will be able to reduce the CBC background energy densities by a factor of about 2, 10 and 200 from the BNS, BH-NS and BBH population respectively through a subtraction scheme. The total residual foreground is overwhelmingly due to sub-threshold BNS merger events and is insensitive to rate evolutionary histories and BH mass distributions. The possibility that $r_0$ for BBH could be much higher than the value used here, e.g., $r_0=0.36 \hspace{0.5mm} \rm{Mpc}^{-3} \hspace{0.5mm} \rm{Myr}^{-1}$ found in \citet{Bulik11_BBH}, does not significantly change the level of such a residual foreground. Additionally the contribution from a possible population of dynamically formed BBHs to such a residual foreground is negligible, as ET will be able to detect these sources out to much larger distances than the field population of BBHs considered in this work due to significantly higher chirp masses \citep{Sadowski08}.

\begin{figure}
\includegraphics[width=0.5\textwidth]{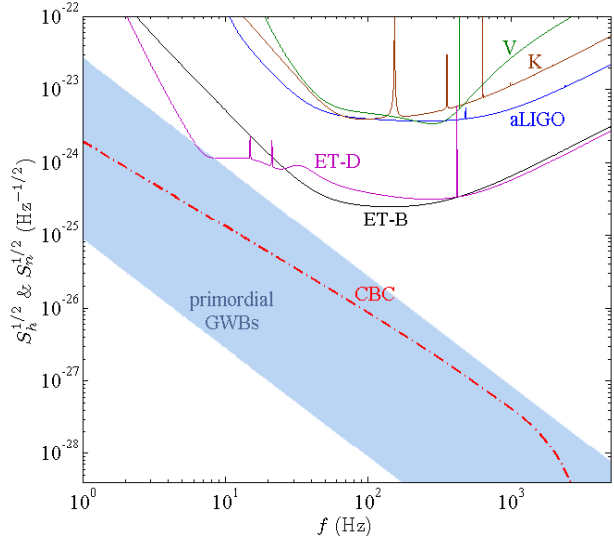}
\caption{The spectral densities, $S_{h}(f)$, of the residual foreground formed by sub-threshold (for ET) CBC events and some putative primordial GWB from the very early Universe -- the shaded region is encompassed by two flat energy spectra $\Omega_{0}^{\rm{UP}} = 10^{-9}$ and $\Omega_{0}^{\rm{LOW}} = 10^{-14}$ (note that they are not upper or lower limit, see text for details), are compared against noise power spectral densities, $S_{n}(f)$, of second (aLIGO, K -- KAGRA, V -- advanced Virgo) and third generation (two possible configurations for ET, ET-B and ET-D, are considered) ground-based GW interferometers.}
\label{ShSnComp}
\end{figure}

Figure \ref{ShSnComp} compares the noise power spectral densities of future terrestrial detectors with the spectral densities of the CBC residual foreground and a range of primordial GWBs from the very early Universe which could be described by a flat energy spectrum in the frequency band of ground-based interferometers. Examples of such primordial GWBs include inflationary, cosmic strings, and pre-Big-Bang models \citep[see Figure 2 in][and references therein for deteails]{LIGO-SGWB-limit}. Considering significant uncertainties associated with model predictions, we show a shaded region formed by $\Omega_{0}^{\rm{UP}} = 10^{-9}$ and $\Omega_{0}^{\rm{LOW}} = 10^{-14}$ with $\Omega_{0}^{\rm{UP}}$ corresponding to the upper limit achievable by aLIGO (a level that could be reached or surpassed in cosmic strings and pre-Big-Bang models) and $\Omega_{0}^{\rm{LOW}}$ for the likely level of inflationary GWBs (which is below the ET stochastic sensitivity). Figure \ref{ShSnComp} implies that, without considering other types of AGWB sources, the contribution to a foreground from sub-threshold CBC events could be a challenging issue for future stochastic searches for primordial GWBs because these signals are beyond the capability of current data analysis methods and always add up to act as an additional ``noise" component in the data. Note that: a) the CBC curve shown in Figure \ref{ShSnComp} only applies to ET, and the foreground level for advanced detectors is $2\times$ higher (as the original pre-subtraction background signal); b) in the case of different coalescence rates and average chirp masses, the total residual foreground level can be estimated in a fashion similar to equation (\ref{SNRscaling}).

\section{Time-frequency properties}
\label{resoveCBC}
The analysis in the previous section is based on the expectation that individual CBC events contributing to a background have different amplitudes in the data due to a distribution over source distances, orientations and sky positions. Rigorously one need to track the number of sources contributing in the relevant frequency intervals. For CBC events, the duty cycle is frequency dependent -- individual signals stay much longer at low frequencies, leading to much smaller duty cycle for increasing frequencies, as pointed out in recent studies \citep[see, e.g.,][]{Rosado11,WuCBC12}.

Practically we want to know the critical frequency, $f_{c}$, above which individual signals do not simultaneously occupy the same frequency interval. Without going into specific details of the calculations, we provide the following relation:

\begin{equation}
\left(\frac{f_{c}}{15\hspace{0.5mm} \rm{Hz}}\right)^{11/3} = \left(\frac{\Delta f}{1\hspace{0.5mm} \rm{Hz}}\right)\hspace{0.5mm} \left(\frac{r_0}{1\hspace{0.5mm} \rm{Mpc}^{-3}\rm{Myr}^{-1}}\right)\hspace{0.5mm} \left(\frac{M_c}{1\hspace{0.5mm} M_{\odot}}\right)^{-5/3}
\label{fDC},
\end{equation}
where $\Delta f$ is the size of frequency interval relevant to the analysis. We consider a reference value of 1 Hz for $\Delta f$, while in practice it could be as small as the frequency resolution of an experiment. For ground-based interferometers such a resolution is given by $1/\Delta T$, where $\Delta T$ is the time duration of short data segments which are used in cross correlation analysis and is typically of orders of seconds \citep{AllenGWB99,LSC12SGWB}. In equation (\ref{fDC}) we neglect the effect of mass distribution without loss of generality.

Once we know the critical frequency $f_{c}$, the \textit{duty cycle function} $\xi (f, \Delta f)$ is exclusively determined through:
\begin{equation}
\xi (f, \Delta f) = \left(\frac{\Delta f}{1\hspace{0.5mm} \rm{Hz}}\right) \left(\frac{f}{f_{c}}\right)^{-11/3}
\label{DCfdf}.
\end{equation}
Note that: a) the duty cycle function is comparable to the overlap function in \citet{Rosado11} and the duty cycle parameter in \citet{WuCBC12}; b) the contribution of post-inspiral emission to $\xi (f, \Delta f)$ is negligible due to much shorter durations; c) the above two convenient relations are applicable to three CBC populations up to a few hundreds Hz.

\begin{figure}
\includegraphics[width=0.48\textwidth]{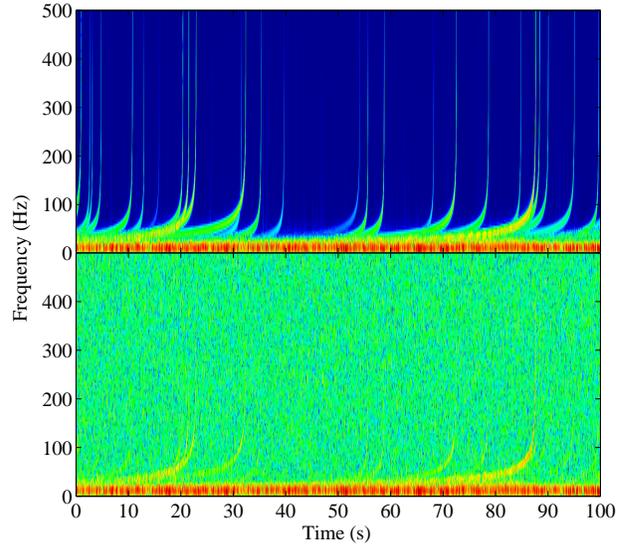}
\caption{The spectrograms for a simulated BNS background signal (upper panel) and for the same background signal plus an arbitrary amount of Gaussian white noise (lower panel). In the simulation we increase the coalescence rate by a factor of 50 and scale down the final amplitude by the same factor. The colors or intensities in the spectrogram effectively tells the relative signal/noise power spectral densities at given time and frequency instances. The lower panel is only for illustration and should not be related to actual detection prospects.}
\label{SpectroBNS}
\end{figure}

The physical meaning of the duty cycle function is the (statistically average) number of intersections of the tracks of individual inspirals in time-frequency plane at a given frequency interval. We demonstrate this in the upper panel of Figure \ref{SpectroBNS} by plotting the spectrogram of a simulated time series of BNS inspirals up to $z=6$. This allows one to visualize the unique time-frequency properties of the BNS background: a) above a few tens Hz, individual \textit{chirps} are only occasionally present, separated in both time and frequency domain; b) a large number of overlapping signals below $\sim 20$ Hz create a continuous and stochastic background, with the colors showing the ``redness'' of the background. The simulation follows the same procedure as described in \citet{Tania_ETMDC} except that we increase $r_0$ by a factor of 50 and scale down the final amplitude by the same factor.

For illustration we add an arbitrary amount of Gaussian white noise to the simulated background, and the spectrogram is shown in the lower panel of Figure \ref{SpectroBNS}. As most of the signals are deeply buried in detector noise, one can only resolve the strongest events which are above the detection threshold, just like a few ``chirping" structures apparent in the noisy spectrogram. Note that the comparison of signals and noise in this plot are not representative of the actual detection prospects since the added noise is not comparable to (and obviously weaker than) the instrumental noise of current and future ground-based detectors.

The conclusion from equation (\ref{fDC}) and Figure \ref{SpectroBNS} is that at each 1 Hz frequency bin above 15 Hz, we will be observing individual BNS inspirals rather than a background. The equivalent critical frequency is of order 4 (2) Hz for the BH-NS (BBH) population. Furthermore, when an array of detectors with moderate angular resolution is used to observe the CBC background, the background must be considered as a set of discrete transient sources randomly distributed across the sky. However, this does not necessarily mean the GW emission due to various CBC populations as a whole can not be detected as a background by the standard cross correlation method (as we have pointed out in the beginning of subsection \ref{SNRresults}). In practice, the underlying numerous individual signals are averaged when cross-correlating data of length from months to years, and it is possible to recover the theoretical expected spectral density as demonstrated in a recent mock data challenge study for ET \citep{Tania_ETMDC}. In the same way, sub-threshold transient signals remain in the data as an additional noise component which could obscure the primordial GWBs and other AGWBs.

\section{Conclusions}
\label{Conclusions}
In this paper we first reviewed the formalism of the calculation of $\Omega_{\rm{GW}}(f)$ and developed a practical model for AGWBs -- a power law energy spectrum ${\rm{d}}E_{\rm{GW}}/{\rm{d}}f \sim f^{\alpha-1}$ naturally leads to $\Omega_{\rm{GW}}(f) = \Omega_{\alpha} f^{\alpha}$ where $\Omega_{\alpha}$ depends almost exclusively on the local rate density $r_0$ and total amount of radiated GW energy $\Delta E_{\rm{GW}}$. Such a model allows one to quickly evaluate uncertainties in estimates of the background strength and the associated detectability.

We have provided updated estimates of the spectral properties of the CBC background formed by populations of BNS, BH-NS and BBH systems. By systematically investigating effects of CSFRs, delay times, NS/BH mass distributions, and using up-to-date analytical complete waveforms including PN amplitude corrections, we showed that:\newline
1) Effects of CSFRs and delay times are linear below 100 Hz and can be represented by a single parameter $J_{2/3}$ with an uncertainty $\sim2$;\newline
2) PN effects cause a small reduction of $\Omega_{\rm{GW}}(f)$ from a $f^{2/3}$ power law function above a few tens Hz;\newline
3) Below 100 Hz, $\Omega_{\rm{GW}}(f)$ can be approximated by a $f^{2/3}$ power law function, with the magnitude determined by only three parameters -- the local coalescence rates $r_0$, the average chirp mass $\langle M_c^{5/3} \rangle$ plus $J_{2/3}$. In particular, within this frequency range $\Omega_{\rm{GW}}(f)$ does not depend on chirp mass distributions. This finding, which was also obtained independently in a recent study using a population-synthesis approach \citep{Kowalska12}, is important for parameter estimation of this background;\newline
4) A variety of features at high frequencies ($\gtrsim 200$ Hz), e.g., different peak frequencies and widths of $\Omega_{\rm{GW}}(f)$, are expected from different CSFRs, delay times, and mass distributions. Measurements of the peaks will be rewarding although challenging due to the small contribution (less than $1\%$) to $\mathrm{S/N}$ by the high frequency signal;\newline
5) The post-merger emission of BNS coalescences could considerably enhance the peak of the BNS background at around 1-2 kHz, but will not alter the background spectrum below 300 Hz. While this contribution to a GWB may be too weak to be detectable even for ET, the latter fact is advantageous for parameter ($r_0$ and $\langle M_c^{5/3} \rangle$) estimation by measuring only the low-frequency power law spectrum.

Using updated estimates of $\Omega_{\rm{GW}}(f)$, we revisited the issue on the detectability of this background signal. Assuming a detection target of the total background contributed by three CBC populations for a worldwide network of advanced detectors, we showed in Figure \ref{SNR3r0} the accessible ``rate space" of the local coalescence rates $r_0$ (in $\rm{Mpc}^{-3} \hspace{0.5mm} \rm{Myr}^{-1}$, with the value of BH-NS fixed at 0.03), implying:\newline
1) A combination of a BNS population at the realistic rate of $r_0 = 1$ and a BBH population at a rate of $r_0 = 0.1$ will give rise to a detectable background signal;\newline
2) Either a BNS rate of $r_0 = 2.7$ or a BBH $r_0 = 0.16$ will be necessary for detection, when BBH or BNS has very low coalescence rate (note that the chosen $r_0$ for BH-NS ensures a negligible contribution).\newline
In both cases, recent optimistic rate estimates for BBHs provide interesting detection prospects for a CBC background. The above quoted values are for optimally combining a network of 5 advanced detectors. Such an optimal combination gives $30\%$ improvement in detectability over aLIGO H-L. This is way below the common expectation \citep{WuCBC12,Kowalska12} that such a network could perform as well as two co-located and co-aligned aLIGO detectors, which gives a 3-fold improvement on H-L. In the latter case our results are consistent with those presented in \citet{WuCBC12}, showing it is likely that at the realistic rate a BNS background may be detected within one year observation using two co-located aLIGO interferometers.

We emphasize that the somewhat ``disappointing" performance of a network of detectors is due to effects of the overlap reduction functions for the current configurations of the advanced detector array -- the large separations between pairs of detectors, of the orders of $10^4$ km (except H-L, 3000 km), result in very modest correlation of background signals above 20 Hz (50 Hz for H-L). This further implies that stochastic background searches can benefit significantly from a pair of closely spaced detectors, with separation chosen to be both within one reduced wavelength (about 300 km for 150 Hz) and relatively large to ensure that their noise sources are largely uncorrelated.

We found that $99\%$ of the $\mathrm{S/N}$ can be obtained by considering only the contribution up to 50 Hz (aLIGO H-L) or at most 100 Hz (KAGRA-advanced Virgo). Two main implications for advanced detectors are:\newline
1) Only the low frequency part is important for detection;\newline
2) Improvement on the sensitivity below 50 Hz is beneficial for detection.\newline
We conclude that a simple power law model as given by equation (\ref{omegCBC0}) with an upper frequency cutoff of 50-100 Hz is sufficient for background searches. Since the model is generalized to three CBC populations and only requires three parameters, it could prove useful to constrain or estimate these parameters with future stochastic searches -- particularly one can marginalize over a uniform distributed $J_{2/3}$ to obtain confidence levels of $r_0$ and $\langle M_c^{5/3} \rangle$. In addition, our generalized model can also be used to identify the relative contribution from different populations in the case of a likely detection of the CBC background. This will further require combination of stochastic background measurements with CBC single event detections \citep{VukPE12}. Regarding the above point 2), we specifically showed that for the CBC background the aLIGO tuning configuration offering the best low-frequency sensitivity (which is optimized for BBH inspiral searches) will provide a $50\%$ enhancement in the achievable $\mathrm{S/N}$ against the standard sensitivity (zero detuning with high laser power), and such an improvement is even better than that due to an optimal combination of the currently proposed detector network (which comprises 5 advanced detectors). We further compared the sensitivities of stochastic searches using different aLIGO tuning options to generic power law GWB models in Figure \ref{MinOmegaLIGO}. The results show that aLIGO H-L will be able to detect a GWB with a $\mathrm{S/N}$ above 3 for $\Omega_{0} \geqslant 1.87 \times 10^{-9}$ (assuming a flat energy spectrum) with one year observation at the standard sensitivity, and this limit could be reduced down to $1.24 \times 10^{-9}$ using the optimal BH-BH option.

For third generation detectors like ET, the background will be easily detectable, with a $\mathrm{S/N}$ from tens up to hundreds contributed by individual populations. The high achievable $\mathrm{S/N}$ will open up the possibilities to: a) enable different populations to be disentangled; b) measure PN effects; c) probe mass distributions and rate evolutionary histories by measuring the (high-frequency) peaks of the background energy spectra. To gain more insights about how these information can be extracted from background measurements, models presented in this study can be further improved in the following ways:\newline
1) Contribution from possible populations of dynamically formed BBHs and/or binaries involved with one or two intermediate-mass BHs \citep[see][and references therein for details]{lsc_rate10} should be considered. Due to significantly higher masses of such systems, their contribution could peak at a few tens Hz and might affect the power law relation for the three normal CBC populations;\newline
2) More accurate complete waveforms are required. In this regard, the three types of BH-NS waveforms corresponding to different merger processes \citep{BH-NS_review11} are of particular interest and can be used to investigate how the information of NS equation of state is encoded in the background signal.

We demonstrated that ET could potentially reduce the contributions to a GWB from the BNS, BH-NS and BBH populations respectively by a factor of 2, 10 and 200 through the subtraction of individually detectable events but there is a strong residual foreground dominated by sub-threshold BNS merger events. Such a foreground, at the level of $\Omega_{\rm{GW}}\sim 10^{-10}$ in the (1--500) Hz frequency range, can hardly be removed and should be considered in future terrestrial searches of primordial GWBs and other AGWBs.

We finally discussed the unique properties of the CBC background -- well defined continuously rising tones, localized directions and well defined average spectral density. These have not so far been fully exploited by stochastic background searches. We believe that new algorithms could exploit these properties to go beyond the standard cross correlation limit that applies only to true stochastic backgrounds.

\section*{Acknowledgments}
We gratefully acknowledge Vuk Mandic for a careful reading of the manuscript and insightful suggestions which have led to some valuable amendments. We also thank Luciano Rezzolla for important feedback on the BNS waveforms used in this study. We are grateful to Tania Regimbau, Eric Thrane, Yanbei Chen and Linqing Wen for useful discussions, and also to the referee for helpful comments. The authors acknowledge iVEC at UWA for the use of computing facilities (Xanax and Valium) in the Monte-Carlo simulation. X-JZ thanks Yun Chen, Xi-Long Fan, Yi-Ming Hu and Zhong-Yang Zhang for helpful discussions. X-JZ acknowledges the support of an University Postgraduate Award at UWA. EJH was supported by an UWA Research Fellowship. Z-HZ was supported by the National Natural Science Foundation of China under the Distinguished Young Scholar Grant 10825313 and the Ministry of Science and Technology National Basic Science Program (Project 973) under Grant No. 2012CB821804.

\bibliography{Zhu2012a}

\bsp \label{lastpage}
\end{document}